\documentclass[useAMS,usenatbib]{mn2e}
\newcommand{\mnras}{MNRAS}

\newcommand{\nat}{Nature}
\newcommand{\araa}{ARAA}
\newcommand{\aj}{AJ}
\newcommand{\apj}{ApJ}
\newcommand{\apjl}{ApJ}
\newcommand{\apjs}{ApJSupp}
\newcommand{\aap}{A\&A}

\newcommand{\apss}{Ap\&SS}

\newcommand{\actaa}{AcA}
\usepackage{graphicx}
\usepackage{rotating}
\usepackage{color}

\title{A highly unequal-mass eclipsing M-dwarf binary in the WFCAM Transit Survey.}
\author[S.V.Nefs]{S.V. Nefs$^{1,}$\thanks{E-mail:nefs@strw.leidenuniv.nl},J.L. Birkby$^{1}$,I.A.G Snellen$^{1}$, S.T. Hodgkin$^{2}$, B.M. Sip{\H o}cz$^{3}$, G. Kov{\'a}cs$^{2}$,\and D. Mislis$^{2}$, D.J. Pinfield$^{3}$ and E. L. Martin$^{4}$.\\
$^{1}$Leiden Observatory, Leiden University, Niels Bohrweg 2,2333 CA Leiden, The Netherlands\\
$^{2}$Institute of Astronomy, University of Cambridge, Madingley Road, Cambridge, CB3 0HA, UK\\
$^{3}$Centre for Astrophysics Research, University of Hertfordshire, College Lane, Hatfield, Hertfordshire AL10 9AB\\
$^{4}$Centro de Astrobiologia (CSIC/INTA), Carretera Ajalvir km 4, 28850 Torrejon de Ardoz, Madrid, Spain\\}
\begin{document}

\date{Accepted March 4}

\pagerange{\pageref{firstpage}--\pageref{lastpage}} \pubyear{}

\maketitle

\label{firstpage}
\begin{abstract}
Star formation theory predicts that short-period M-dwarf binaries with highly unequal-mass components are rare. Firstly, the mass ratio of close binary systems is driven to unity due to the secondary preferentially accreting gas with high angular momentum. Secondly, both dynamical decay of multiple systems and interactions with tertiary stars that tighten the binary orbit will eject the lowest mass member. Generally, only the two most massive stars are paired after such interactions, and the frequency of tight unequal-mass binaries is expected to decrease steeply with primary mass. In this paper we present the discovery of a highly unequal-mass eclipsing M-dwarf binary, providing a unique constraint on binary star formation theory and on evolutionary models for low-mass binary stars. The binary is discovered using high-precision infrared light curves from the WFCAM Transit Survey (WTS) and has an orbital period of 2.44 d. We find stellar masses of $M_1= $ 0.53 $\pm $ 0.02 $ M_{\odot}$ and $M_2= $ 0.143 $\pm $ 0.006 $ M_{\odot}$ (mass ratio 0.27), and radii of $R_1= $ 0.51 $\pm $ 0.01 $ R_{\odot}$ and $R_2= $ 0.174 $\pm $ 0.006 $ R_{\odot}$. This puts the companion in a very sparsely sampled and important late M-dwarf mass-regime. Since both stars share the same age and metallicity and straddle the theoretical boundary between fully and partially convective stellar interiors, a comparison can be made to model predictions over a large range of M-dwarf masses using the same model isochrone. Both stars appear to have a slightly inflated radius compared to 1 Gyr model predictions for their masses, but future work is needed to properly account for the effects of star spots on the light curve solution. A significant, subsynchronous, $\sim$2.56 d signal with $\sim$2\% peak-to-peak amplitude is detected in the WFCAM light curve, which we attribute to rotational modulation of cool star spots. We propose that the subsynchronous rotation is either due to a stable star-spot complex at high latitude on the (magnetically active) primary (i.e. differential rotation), or to additional magnetic braking, or to interaction of the binary with a third body or circumbinary disk during its pre-main-sequence phase.

\end{abstract}


\section{Introduction}
Stellar evolution models find it difficult to accurately reproduce the fundamental properties of M-dwarf stars ($M_*<$0.65$M_{\odot}$), which are the most abundant population of stars in the Milky Way (over 70\% in number; Henry et al. 1997). Furthermore, the formation process of binary M-dwarfs and their migration to close orbits are not well-understood (e.g. Goodwin \& Whitworth 2007; Clarke 2012; Nefs et al. 2012). M-dwarfs are important to astrophysics because they help us understand a variety of problems, from the local star formation history, to the shape of the stellar (initial) mass function (e.g. Reid 1999; Gizis, Reid \& Hawley 2002; Bochanski et al. 2007). Moreover, M-dwarfs are now being recognised as prime targets in the hunt for Earth-like exoplanets (e.g. Mart\'in et al. 2005; Nutzman \& Charbonneau 2008), as their small stellar size results in deeper transits and their smaller stellar mass causes larger reflex motions induced by planetary companions. However, the current uncertainty in our understanding of M-dwarf formation and evolution means that the parameters of their planets, which scale with those of the host star, can not be determined to high accuracy, limiting a detailed characterisation of their compositions and atmospheres.

Hydrodynamical simulations (e.g. Bate, Bonnell \& Bromm 2002) show that when a star-forming cloud collapses, it fragments into low-number multiple systems, which can be broken up by dynamical interactions (e.g. Goodwin et al. 2007). It is predicted that short-period M-dwarf binaries with significantly unequal-mass components should be very rare. Firstly, this hypothesis is motivated by the Bate, Bonnell \& Bromm (2002) simulations that suggest that infalling gas with high angular momentum is preferentially accreted onto the lower-mass component of a binary during formation, driving the mass-ratio to unity (see also Bate \& Bonnell 1997; Bate 2000). Close binaries in particular are expected to experience significant accretion of gas in order to migrate to tighter orbits. The angular momentum of the gas increases as the accretion proceeds and is higher for closer binary systems, which accrete more gas relative to their initial masses than wider binaries. Whereas gas with low angular momentum is mainly accreted by the primary because it essentially falls straight onto the centre of mass of an unequal-mass system, high angular momentum gas falls in further away, closer to the orbital radius of the secondary, and need not to gain as much momentum to be captured by the companion (e.g. Bate \& Bonnell 1997). 

An enhanced preference of equal mass ratios is expected especially for tight low-mass binaries. This is because dynamical decay of multiple systems and exchange interactions with single stars in the collapsing cloud, which increase the binary mass ratio and at the same time tighten the binary orbit, are biased against the lowest mass stars. In dynamical decay typically the least massive component is ejected on a short timescale, due to the instability of multiple star systems (e.g. Anosova 1986). Furthermore, in dynamic exchange interactions the lowest mass star is removed and replaced by the higher mass intruder. This means that only the two most massive stars survive an interaction, indicating that both the occurrence frequency of binary and multiple systems and the number of close and unequal-mass binaries steeply decrease with decreasing primary mass. Therefore each of the mechanisms involved in producing close and low-mass binaries favors the production of equal-mass systems and highly unequal, short-period, M-dwarf binary systems should thus be rare.

There is observational evidence for this hypothesis. Wisniewski et al. (2012) recently propose a lack of unequal-mass stellar binaries at periods shorter than $\sim$100 d, combining current results from radial velocity, transit and imaging studies. Delfosse et al. (2004) find that M-dwarf binaries with orbital period $P_{orb}<50$ d possess a mass-ratio distribution which is peaked around 1 ('twins'), whereas wider binaries have a flat distribution. Clarke, Blake \& Knapp (2012) find a frequency of close M-dwarf binary stars with a separation of less than 0.4 AU of only 3.0$_{-0.9}^{+0.6}$\%, and argue that the frequency of $<$0.4 AU binaries is decreasing from 26\% at 10$M_{\odot}$ to 1\% at 0.1$M_{\odot}$ (see e.g. Lada (2006) and Raghavan et al. 2010). Their findings suggest that low-mass stars not only have lower multiplicity but their tightest systems are also intrinsically rare. Bouchy et al. (2011) propose that G-type or lower mass stars have stronger disk braking than more massive stars, which causes any short-period low-mass companion to migrate inwards and become engulfed by the primary. 

Detached, double-lined eclipsing binaries (EBs) provide a model-independent method for accurately calibrating the formation and evolution of stars (Andersen et al. 1991; Torres, Andersen \& Gim\'enez 2010). Dynamical measurements of the masses and radii and empirical measurements of the temperatures of M-dwarf stars in eclipsing binary systems, suggest that current models under-predict the radii of M-dwarfs by $5-15\%$, and over-predict their effective temperatures by $3-5\%$ (typically 100-200K; e.g. Lopez-Morales \& Ribas 2005; L\'opez-Morales 2007; Torres, Andersen \& Gim\'enez 2010). These discrepancies are attributed to possible metallicity variations (e.g. Berger et al. 2006), but more likely they are due to strong magnetic fields and spots present on the stars of the observed tidally-locked, short-period M-dwarf binaries (e.g. Mullan \& MacDonald 2001; Ribas 2006; Chabrier, Gallardo \& Baraffe 2007). However, there is no current model that can accurately reproduce all of the observed dynamical measurements.

It is thought that stars with masses $M<0.35M_{\odot}$ converge towards agreement with the current models because they are likely to have fully-convective atmospheres, and thus suffer less from the inflating effects of magnetic inhibition (Kraus et al. 2011). However, even the longest-period, non-synchronised M-dwarf EBs (MEBs) with secondary components in the fully-convective regime (Irwin et al. 2011; Doyle et al. 2011), still show significant radius inflation, despite much lower magnetic fields (see Birkby et al. 2012). For the lowest mass main-sequence M-dwarfs ($0.08M_{\odot}<M<0.2M_{\odot}$), there is even more uncertainty, due to a paucity of model-independent dynamical measurements. Only a few young objects in the Orion Nebula have data (see Irwin et al. 2007; Stassun, Mathieu \& Valenti 2007). Constraints on M-dwarf evolution isochrones are further hindered by the apparent preference for equal mass MEBs. Low-mass ratio MEBs are valuable because their shared age and metallicity allow a more stringent assessment of the stellar model predictions over the wide span of M-dwarf masses. 

In this paper we describe the discovery and characterisation of a main-sequence unequal mass, short-period, detached M-dwarf eclipsing binary system, whose components straddle the fully-convective boundary, and whose secondary star resides in the uncharted $<0.2M_{\odot}$ mass regime. The binary is discovered using high-precision infrared light curves from the WFCAM Transit Survey (WTS; Birkby, Hodgkin \& Pinfield 2011; Birkby et al. 2012; Kovacs et al., submitted). The WTS is an ongoing photometric monitoring campaign that operates as a back-up program running on the 3.8m United Kingdom Infrared Telescope (UKIRT) at Mauna Kea, Hawaii. By observing in the infrared, the WTS is optimised for precision photometry of cool low-mass stars. Its primary objective is to hunt for planets orbiting M-dwarfs by regular monitoring of $\sim$6000 early- to mid M-dwarfs (for J$<$16), but also to characterise the host stars. 

In Section 2 we present the observations and the data reduction of infrared and optical time-series of the eclipses of our binary, WTS 19g-4-02069, and present low- to medium resolution spectroscopy in the optical and in the infrared H-band. In Sections 3 and 4 we characterise the components of WTS 19g-4-02069 using the available data, obtaining individual masses, radii, effective temperatures and activity indicators. In Section 5 we discuss the significance of the binary in the context of current low-mass stellar evolution models. 

\section{Observations \& Data Reduction}

\subsection{WTS J-band time series photometry}
The WTS, in operation since 2007 August, is awarded 200 nights of observing time on UKIRT. The survey uses the UKIRT Wide Field Camera (WFCAM), which has four 2048$\times$2048 18 $\mu$m HgCdTe Rockwell Hawaii-II imaging arrays that each cover 13.65$\times$13.65 arcmin$^2$ on sky (with a pixel resolution of 0.4 arcsec pixel$^{-1}$), and are seperated by 94 per cent of a chip width (Casali et al. 2007). Observations for the WTS are obtained in the J-band (1.25 $\mu$m), which maximises the sensitivity to M-dwarfs with effective temperatures $T_{eff}<$4000K. 

The survey targets four 1.5 deg$^2$ fields centered around 3h, 7h, 17h and 19h in Right Ascension, selected to give both year-round visibility, an optimal number of dwarfs versus giants, relatively low reddening [E(B-V) between 0.057 and 0.234] and reduced contamination by blending stars, by observing close to but outside the galactic plane (galactic latitude $b>$5$^\circ$). The infrared light curves have an average cadence of 15 min. For each field, single deep exposures in the full WFCAM $ZYJHK$ system are also obtained, to aid the photometric identification of M-dwarfs through fitting of the broad band spectral energy distributions (SEDs). The SED for WTS 19g-4-02069, extending from Sloan Digitized Sky Survey (SDSS) $u$ band to infrared Wide Field Infrared Survey Explorer (WISE) 4.6$\mu$m, is shown in Table 1. The observing strategy, pipeline data reduction and WTS light curve generation is extensively described in Birkby et al. (2012) and Kovacs et al., submitted and the interested reader is referred to these two publications for more details.

\begin{figure*}
\begin{center}$
\begin{array}{llllll}
\includegraphics[width=0.5\textwidth]{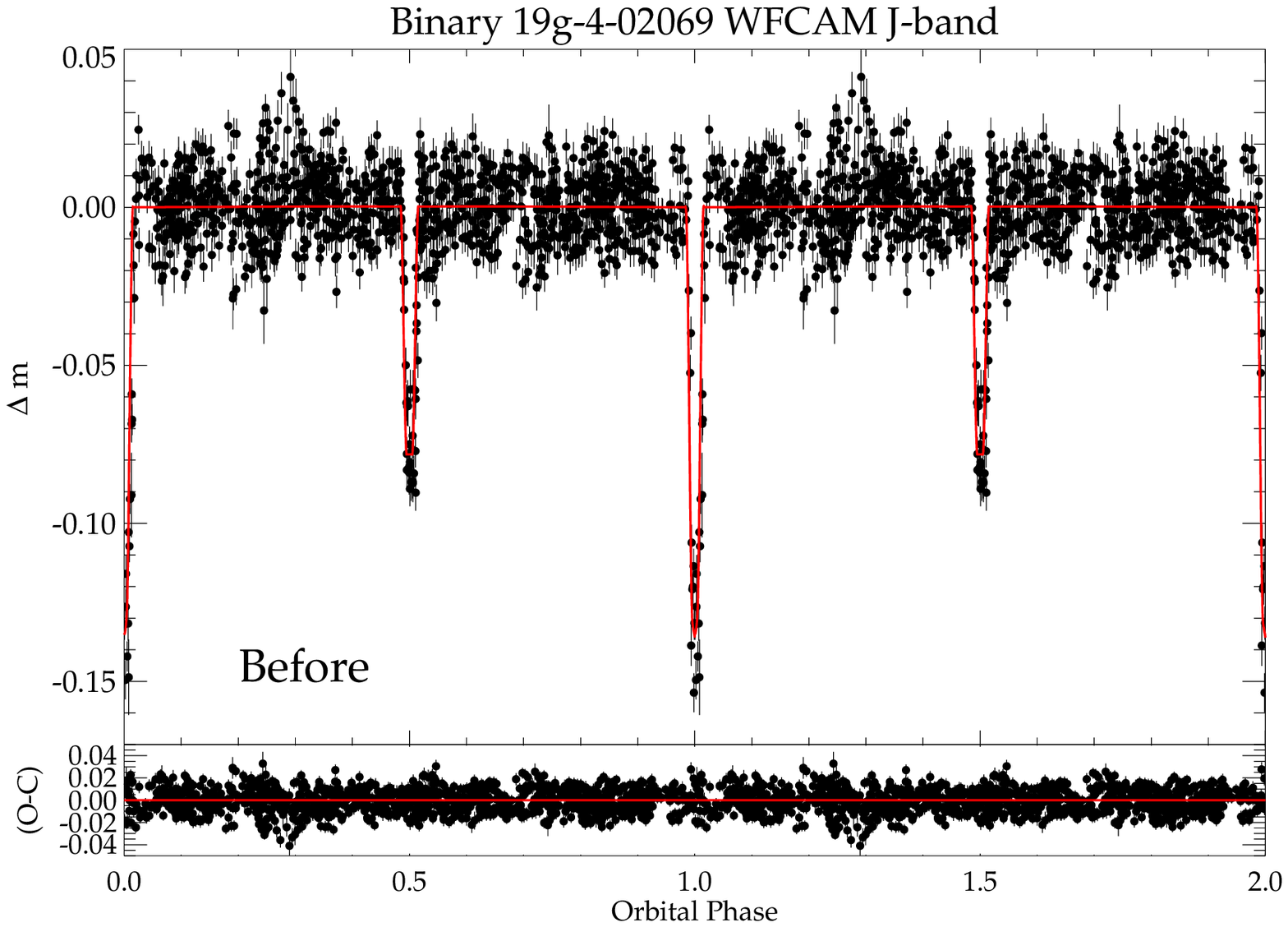} &
\includegraphics[width=0.5\textwidth]{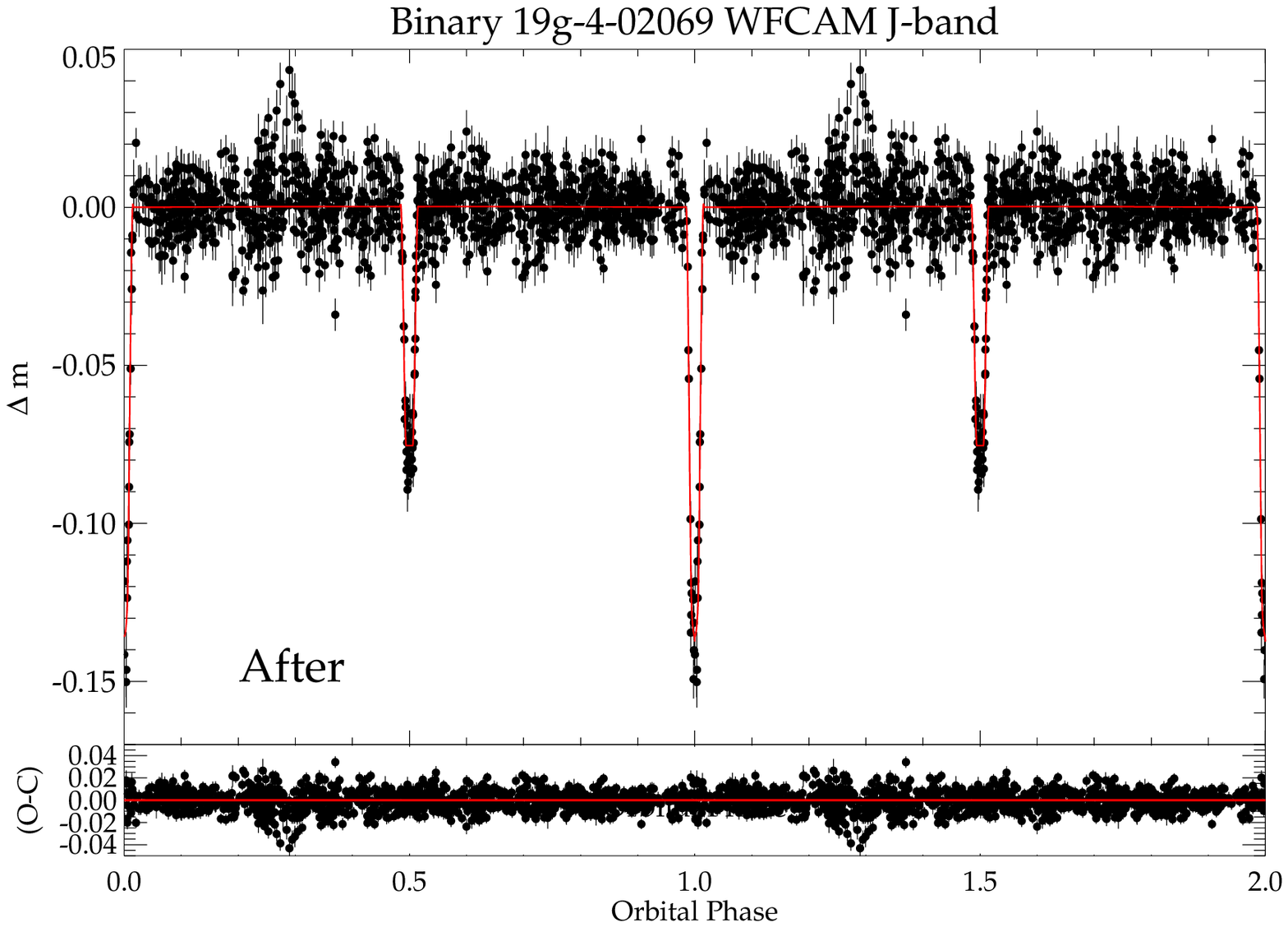} \\
\includegraphics[width=0.5\textwidth]{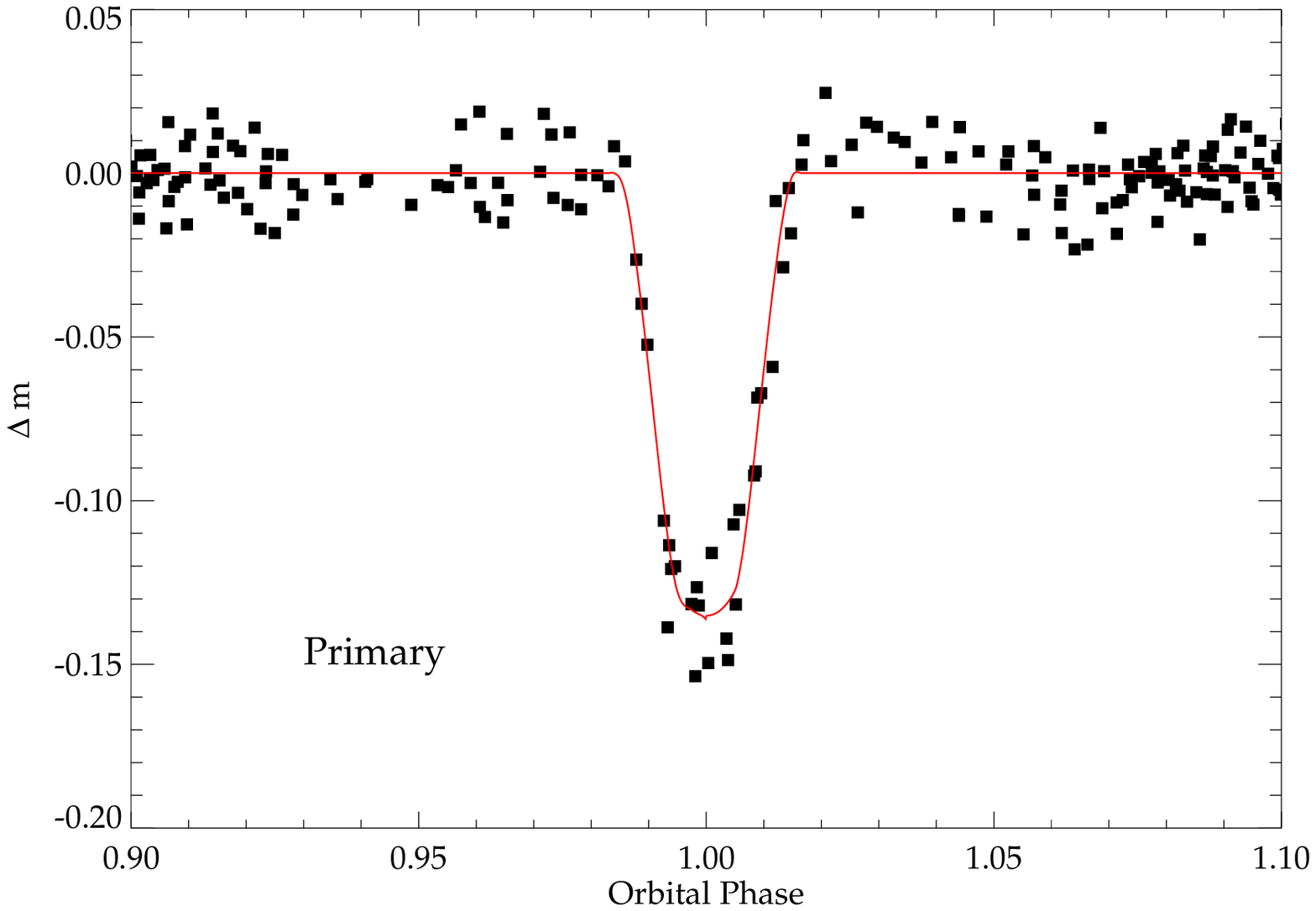} &
\includegraphics[width=0.5\textwidth]{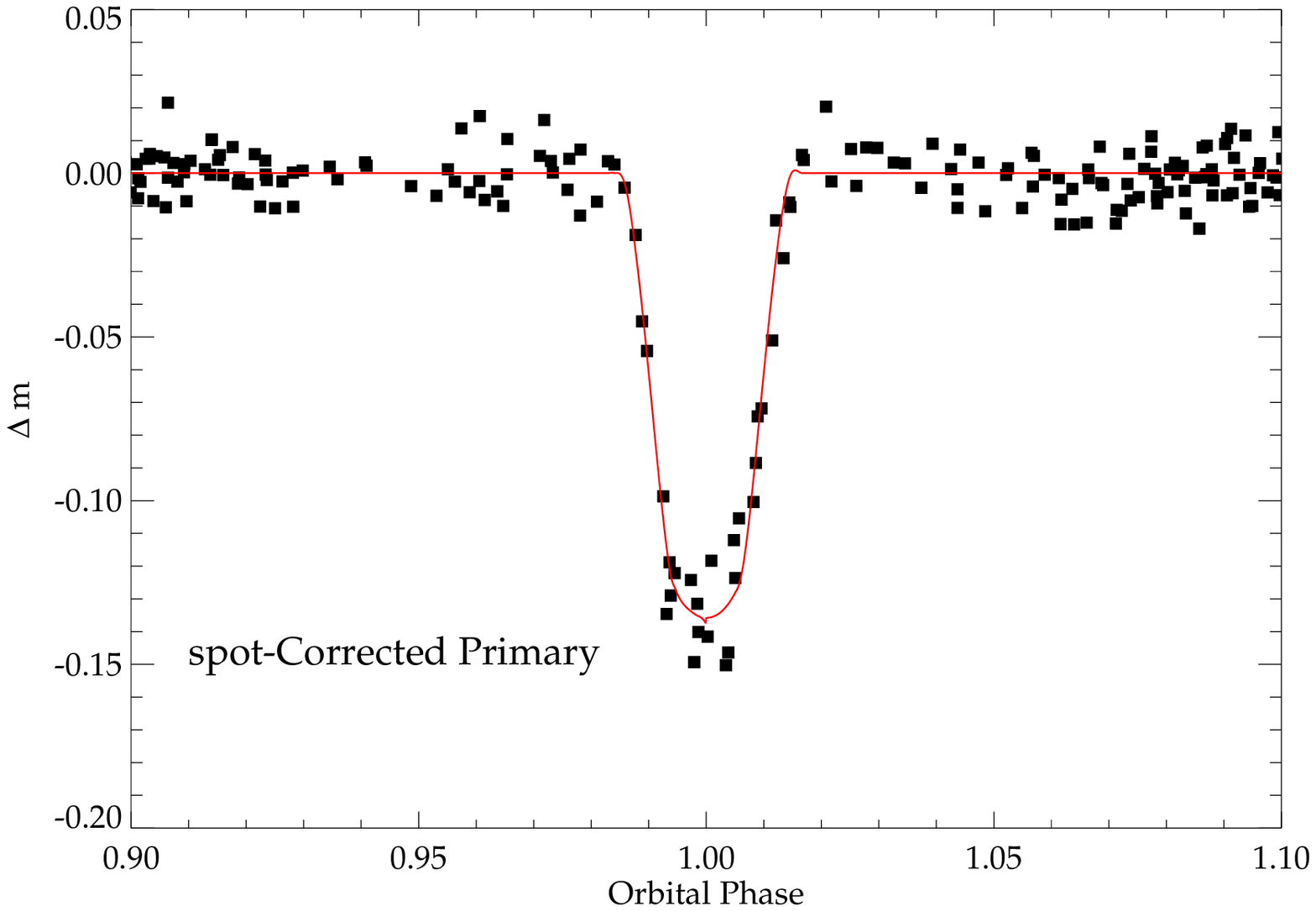} \\
\includegraphics[width=0.5\textwidth]{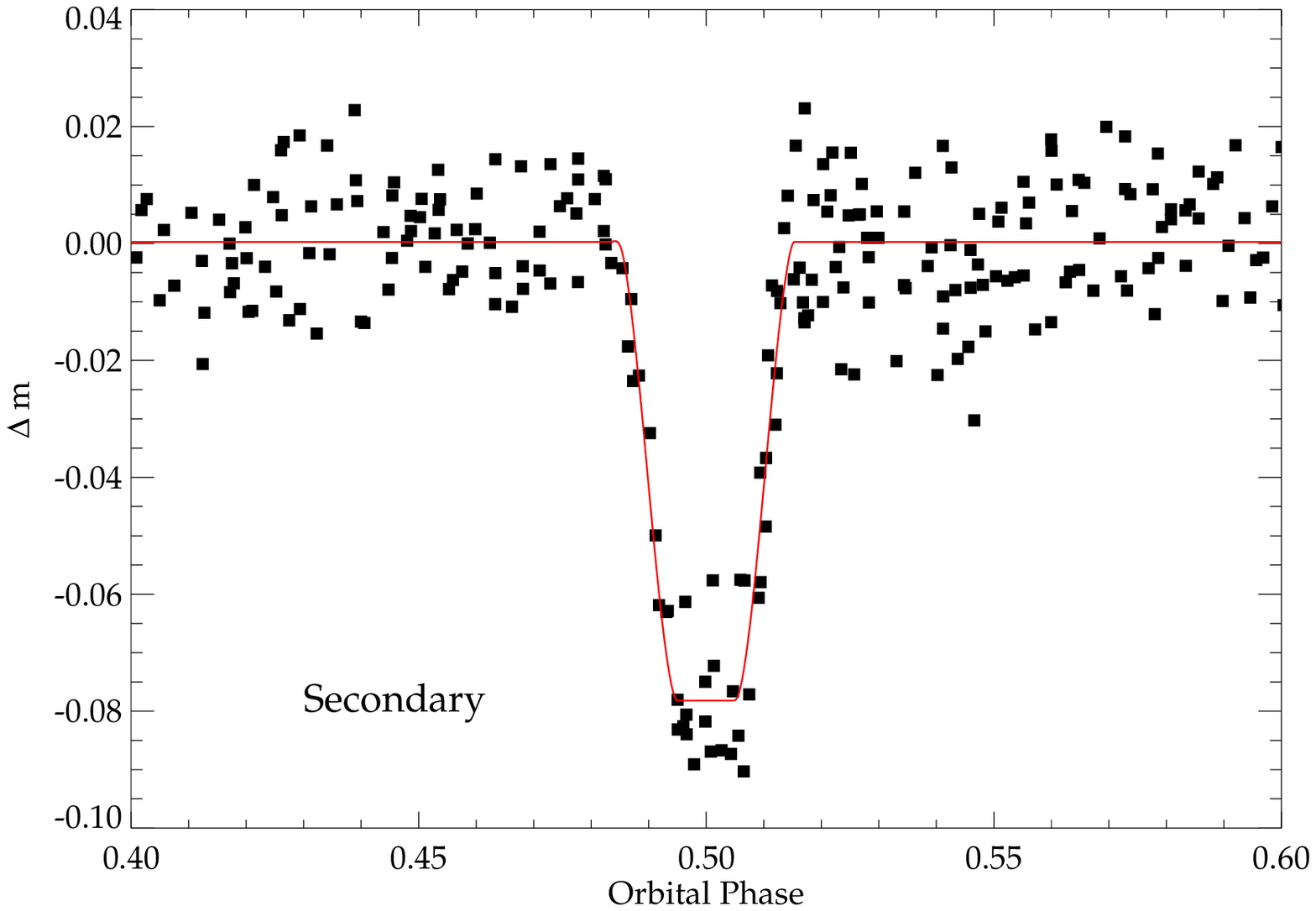} &
\includegraphics[width=0.5\textwidth]{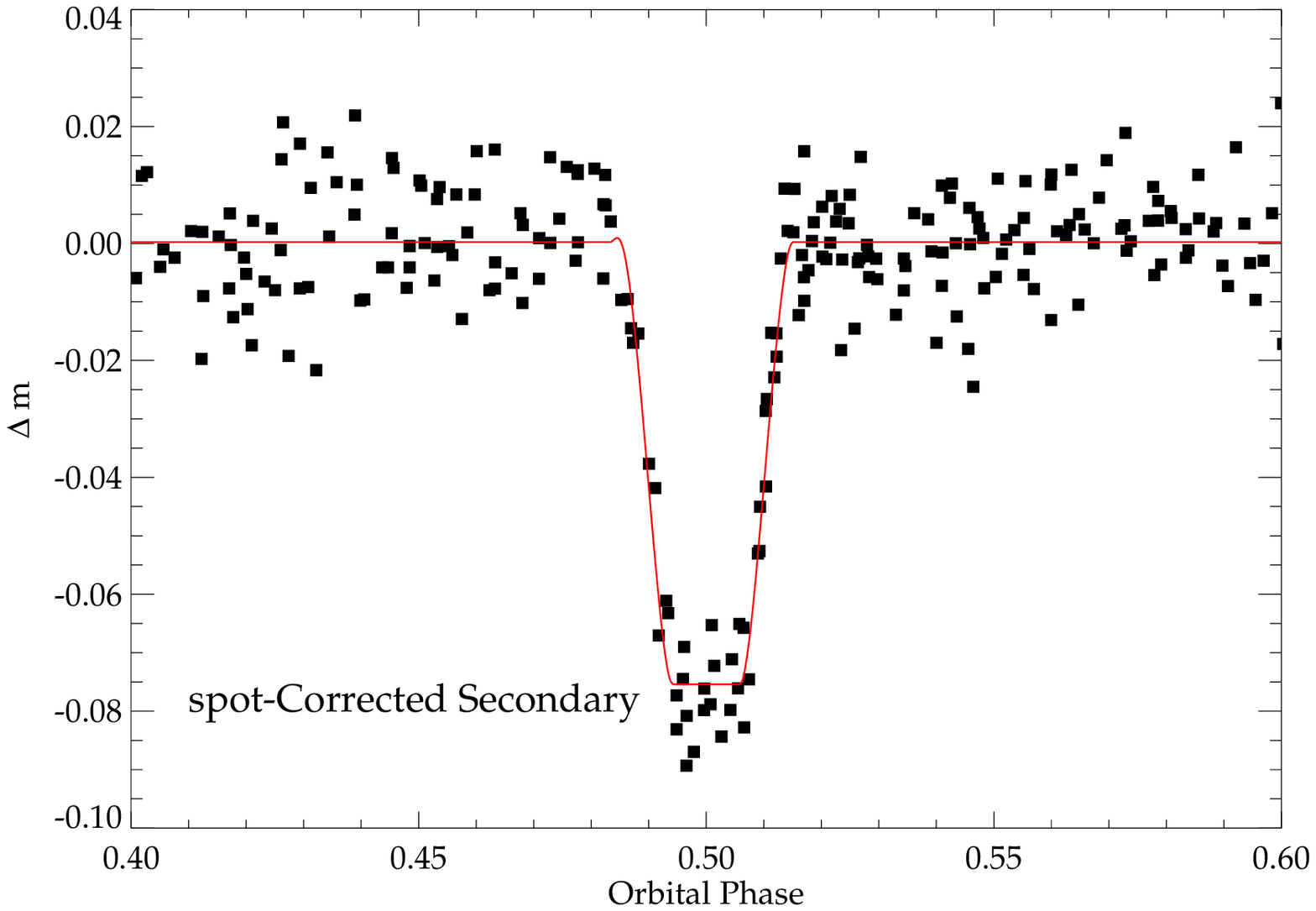} \\
\end{array}$
\end{center}
\caption{\small{The WFCAM J-band discovery light curve of binary WTS 19g-4-02069, before (left upper panel) and after (right upper panel) removal of the rotational signal, which reduces the total light curve rms by $\sim$20\%, and corrects the shape of the eclipses. The data are shown in relative magnitudes $\Delta m$. The best-fit {\sc jktebop} models are overplotted with a solid red line. We also show zoom-ins around the primary and secondary eclipse to illustrate the effect of the correction.}}
\label{Figure1}
\end{figure*}

The subject of this paper is selected from the list of 16 well-sampled detached M-dwarf eclipsing binaries with $J\leq16$ as presented in Birkby et al. (2012) in the 19hr field, the WTS target field which currently has the most extensive observational coverage ($\sim$1200 epochs). An initial source detection is performed using the Box-Least-Squares (BLS) algorithm, {\sc OCCFIT} (Aigrain \& Irwin 2004; Miller et al. 2008). Fitting of the broad band SED of WTS 19g-4-02069 yields a system effective temperature of $T_{eff}\sim$3050K, the lowest of the Birkby et al. (2012) sample, indicating a low-mass eclipsing M-dwarf binary system. The out-of-eclipse root mean square (rms) scatter of WTS 19g-4-02069 is relatively high compared to other stars of similar magnitude ($\sim$12.5mmag per datapoint), whereas $\sim$6mmag is expected. We attribute this to stellar activity (see the discussion in Section 4.1.1). The WTS J band data for WTS 19g-4-02069 are given in Table 2 and shown in Figure \ref{Figure1}, folded on the binary orbital period.
\begin{table}
   \begin{tabular*}{0.5\textwidth}
   {@{\extracolsep{\fill}}lll}
   \hline \hline
   $\alpha_{J2000}$	&19:35:03.55	&\\
   $\delta_{J2000}$	&+36:31:16.49	&\\
   Broad-band SED	&WTS 19g-4-02069	&\\
   \hline
   {\bf{Band name}} &{\bf{Central $\lambda$($\mu$m)}}	&{\bf{magnitude}}\\
   u$^1$ &0.35	&23.594($\pm$3.0)\\
   g	&0.47	&20.262($\pm$0.02)\\
   r	&0.62	&18.866($\pm$0.01)\\
   i	&0.75	&17.277($\pm$0.01)\\
   z	&0.89	&16.335($\pm$0.01)\\
   Z	&0.89	&15.928($\pm$0.005)\\
   Y	&1.03	&15.426($\pm$0.007)\\
   J	&1.25	&14.843($\pm$0.004)\\
   H	&1.63	&14.271($\pm$0.004)\\
   K	&2.20	&13.952($\pm$0.004)\\
   WISE1	&3.4	&13.842($\pm$0.028)\\
   WISE2	&4.6	&13.812($\pm$0.044)\\
   \hline
   \end{tabular*}
\caption{Broad-band spectral energy distribution for the eclipsing M-dwarf binary WTS 19g-4-02069. SDSS u,g,r,i and z magnitudes are quoted in the AB magnitude system, whereas the WFCAM Z,Y,J,H and K magnitudes are in the Vega system. $^1$SDSS $u$ is uncertain because of a red leak (Abazajian et al. 2004). The entries WISE1 and WISE2 refer to the first two wavelength channels of the WISE at 3.4 and 4.6$\mu$m. The source is too faint for detection at 12 and 22$\mu$m.}
\label{generalprop}
\end{table}
\begin{table}
   \begin{tabular*}{0.5\textwidth}
   {@{\extracolsep{\fill}}llll}
   \hline \hline
   $\bf{WFCAM}$	&HJD	&$J_{WTS}$	&$\sigma J_{WTS}$	\\
   J-band		&-2454000	&		&			\\
   			&(days)		&(mag)		&(mag)	\\
   \hline
		&317.808593 &14.8003   &0.0050\\
		&....	&....	&....\\
   $\bf{INT}$	&HJD	&$\Delta m_{iINT}$	&$\sigma m_{iINT}$	\\
   i-band	&	&(mag)		&(mag)			\\
   \hline
   Primary event	&	&	&\\
                        &1403.904630   &0.00146   &0.00455\\
                        &....   &....   &....\\
   Secondary event      &	&	&\\
		        &1409.998380  &-0.00071   &0.00455\\
			&....	&....	&....\\
   \hline
\end{tabular*}
\caption{\small{Photometry for binary WTS 19g-4-02069 showing, from top to bottom, the WTS J-band photometry of WTS 19g-4-02069, and the INT $i$ band data. Quoted magnitudes in the WFCAM system (column 3) can be converted to other photometric systems as described in Hodgkin et al. (2009). The J-band errors $\sigma_j$, are estimated using a noise model including Poisson noise, sky noise, readout noise and errors in the background estimation. (This table is published in full in the online journal and is shown partially here for guidance regarding its form and content).}}
\label{proptable}
\end{table}  
\subsection{INT i'-band follow-up photometry}
We obtain follow-up photometric observations in the Sloan $i$-band on the 2.5m Isaac Newton Telescope (INT) on La Palma, using the Wide Field Camera (WFC), to refine our best-fitting light curve solution from the WFCAM J-band survey data. This imaging camera has a field of view of approximately 34$\times$34 arcmin$^2$ at prime focus, comprised of a mosaic of four 2$\times$4k pixel CCDs, with a resolution of $\sim$0.33 arcsec/pixel. The observations are part of a wider WTS follow-up campaign to confirm planetary transit candidates, between July 18 and August 01 2010, leaving a few windows to observe binary eclipses. We use the WFC in fast readout mode (readout time 28 sec., for 1$\times$1 binning) to observe a full primary eclipse of WTS 19g-4-02069 on the night of July 25 2010 and a full secondary eclipse on the night of July 31 2010. We center the observations around the predicted times of eclipse (based on the J-band data), and allow $\sim$50 min. of observations on either side of the 
predicted eclipses to acquire sufficient baseline. We obtain 82, 90 sec. exposures for the primary event and 52, 200 sec. exposures for the secondary, with out-of-eclipse rms of $\sim$5.6mmag and $\sim$2.2mmag. Error-bars on the data are obtained by assuming a $\chi^2$ value of 1 for the out-of-eclipse parts using the models of Section 4.2. We estimate shot noise errors of $\sim$3.3 mmag (90 s integrations) and $\sim$2.2 mmag (200 s).

\begin{figure*}
\begin{center}$
\begin{array}{ll}
\includegraphics[width=0.5\textwidth]{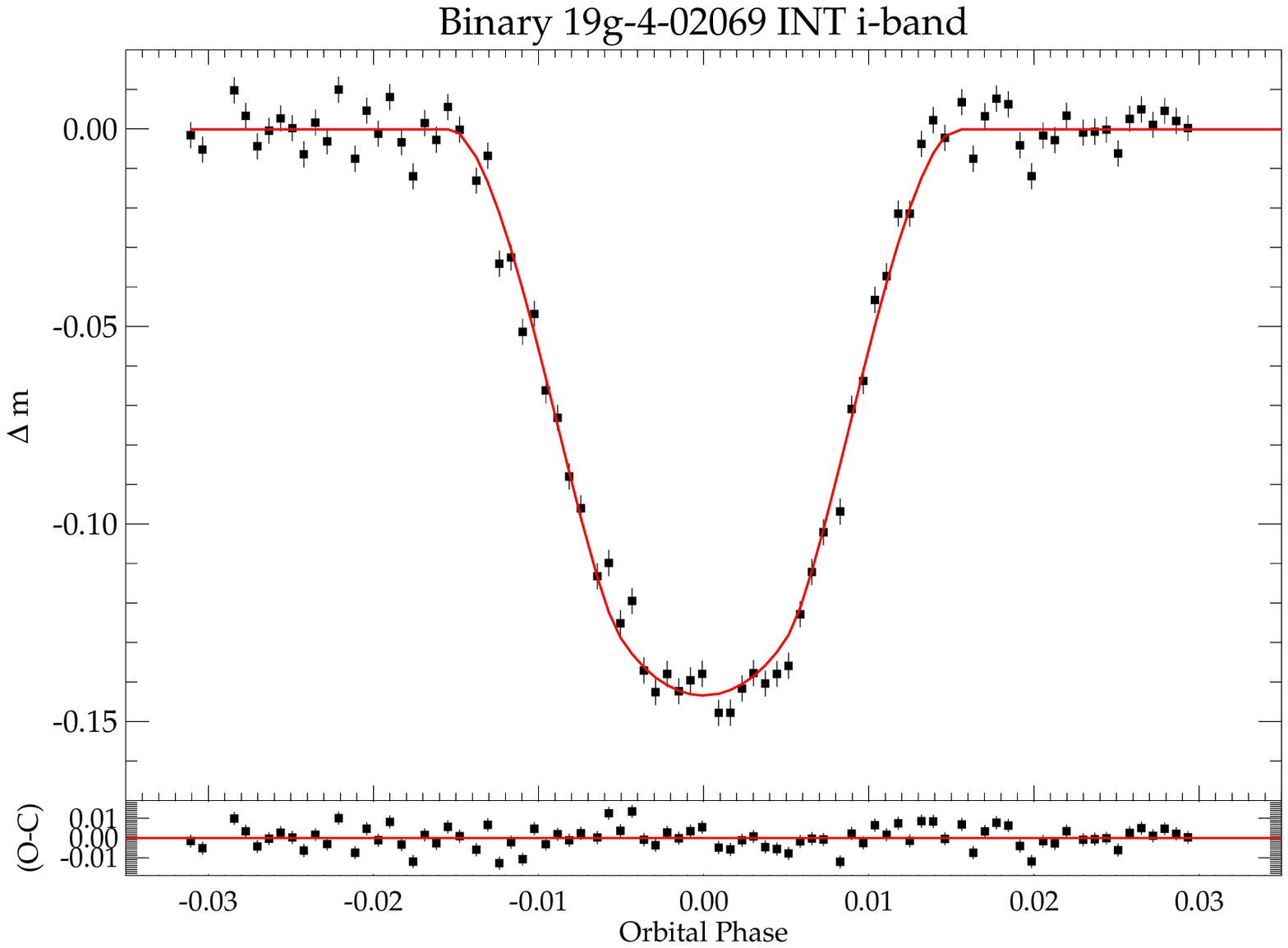} &
\includegraphics[width=0.5\textwidth]{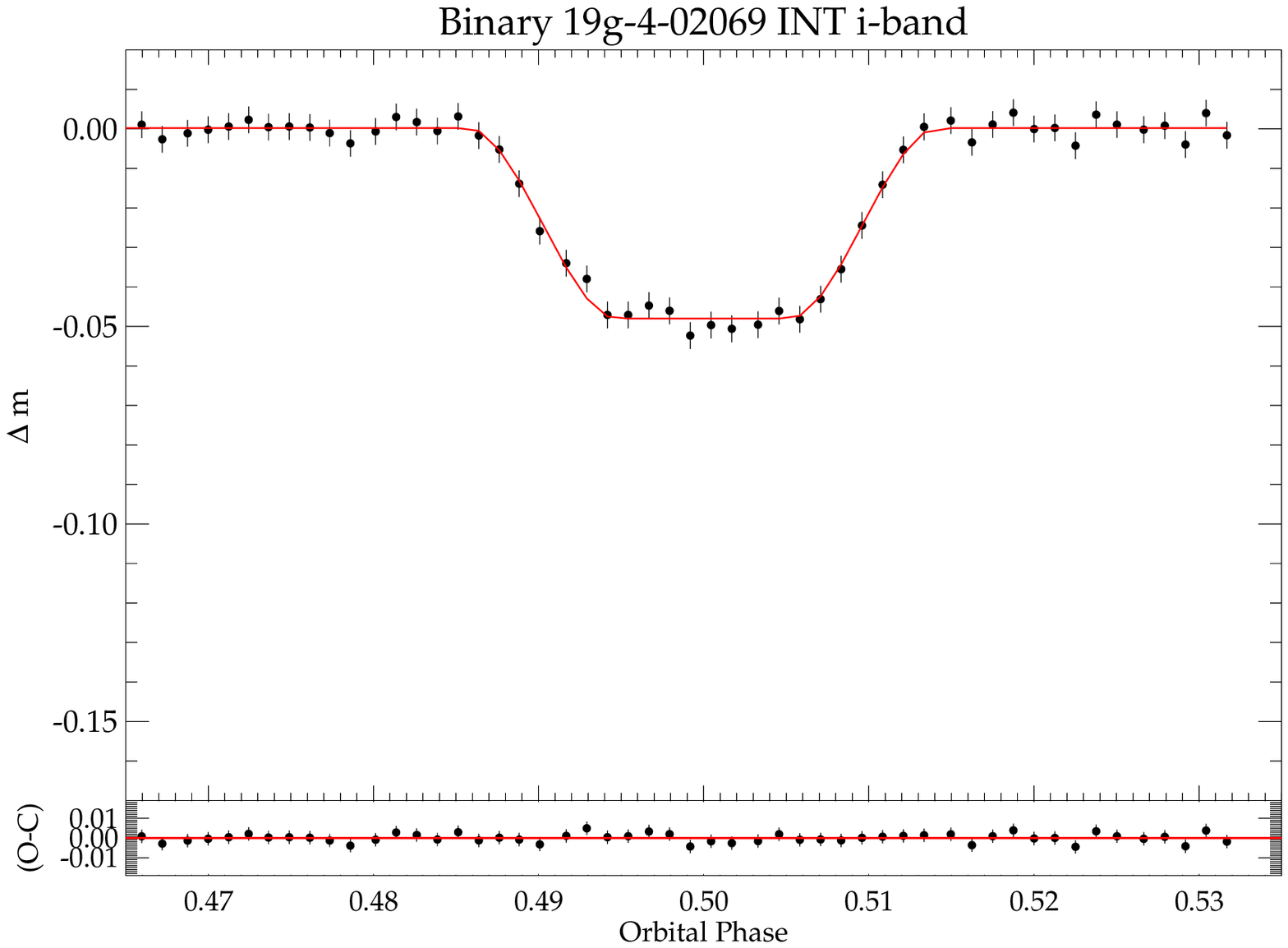} \\
\end{array}$
\end{center}
\caption{The follow-up INT i'-band light curves of the primary (left panel) and the secondary eclipse (right panel) of binary WTS 19g-4-02069. Overplotted in red is the best-fit {\sc jktebop} model.}
\label{Figure2}
\end{figure*}

We reduce the data using custom-built {\sc idl} routines to perform the standard 2-D image processing (i.e. bias subtraction and flat-field division). We remove low-level fringing by subtracting a scaled super sky-frame, which is obtained by median averaging ditthered exposures of a blank field under dark sky conditions. To generate the light curves, we use variable aperture photometry and circular apertures with the {\sc idl} routine {\sc aper}. We estimate the sky background using a 3$\sigma$ clipped median on 30$\times$30 pixel boxes. A master reference light curve is obtained from differential photometry on a set of $\sim$10  bright, nearby, non-saturated, non-blended reference stars, selecting for each reference star the aperture that minimises the out-of-eclipse rms. Airmass dependence is removed by fitting a second order polynomial to the out-of-eclipse data. The INT $i$-band data are presented in Table 2, and shown in Figure \ref{Figure2}.
\subsection{WHT ISIS spectroscopy}
\subsubsection{Low resolution reconnaissance spectrum}
We obtain low-resolution spectroscopy using the William Herschel Telescope (WHT), to confirm the M-dwarf nature of the binary system WTS 19g-4-02069 via measurement of the strengths of gravity sensitive atomic stellar absorption lines, and to estimate the system effective temperature and chromospheric activity. Observations are carried out on the night of July 16, 2010 using the Intermediate dispersion Spectrograph and Imaging System (ISIS). We use the R158R grating, which has a spectral resolution of R$\sim$1200 (1.81 \AA /pixel) at 8500 \AA$ $  and wavelength range of $\sim$6000-9000 \AA, for our 1.0" wide slit, chosen to match the typical seeing conditions of the night. We opt to use ISIS red arm only with the red sensitive RED+ array and not use the dichroic because it can cause systematics and loss of efficiency up to $\sim$10\%. A single 500 sec. spectrum is obtained using this setup at an airmass of $\sim$1.35.

\begin{figure*}
\includegraphics[width=0.87\textwidth]{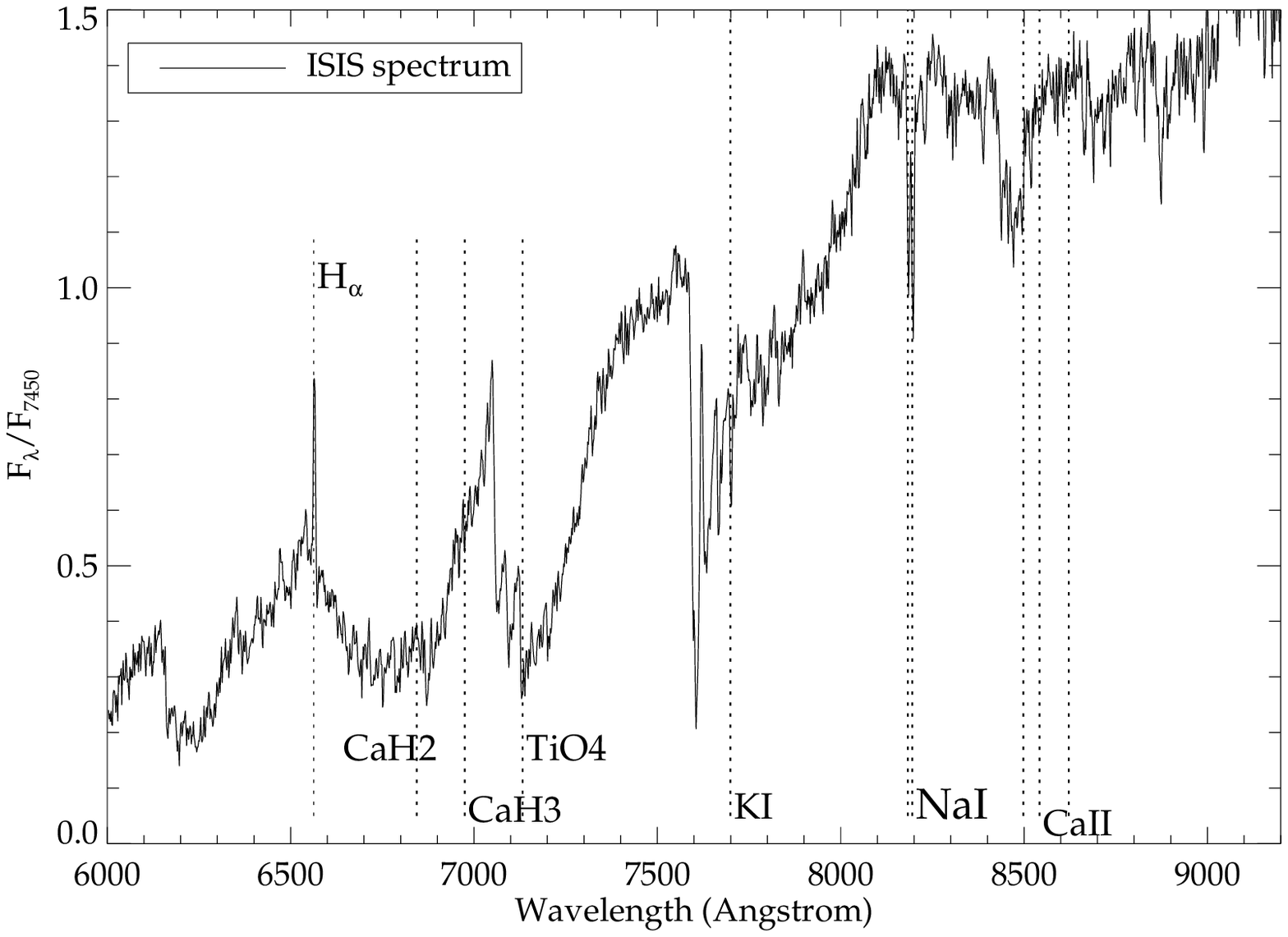}
\includegraphics[width=0.87\textwidth]{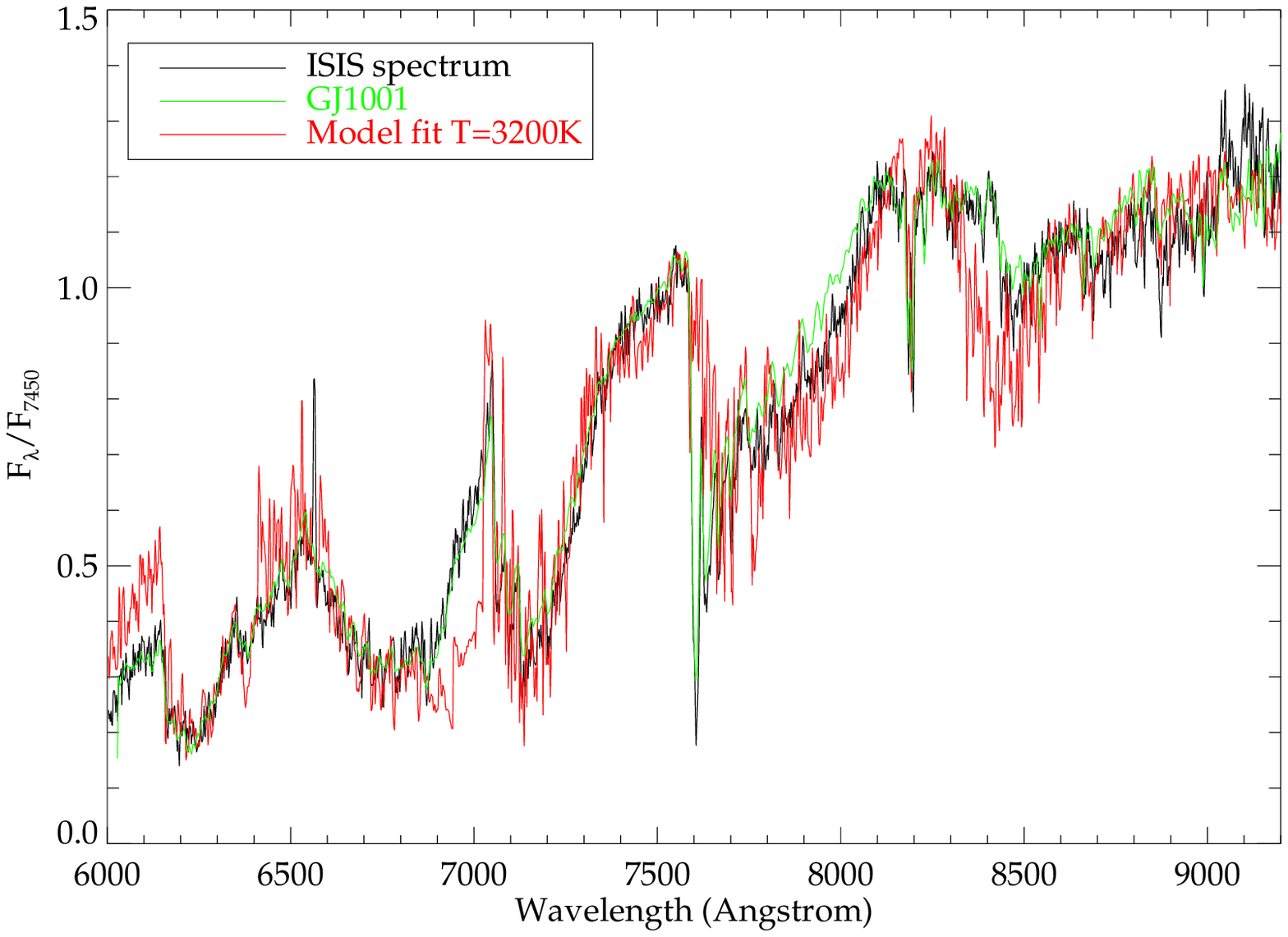}
\caption{\small{Upper panel: The observed low-resolution spectrum of binary WTS 19g-4-02069 obtained with the WHT (solid black line), marking the main molecular and atomic features.The left most vertical dotted line marks $\rm H_{\alpha}$ in emission for WTS 19g-4-02069, suggesting active chromospheres.Lower panel: ISIS spectrum with overplotted the best-fit NexTGen model for $T_{eff}$=3200K (solid red line). The solid green line is an observed template spectrum for GJ1001, a single M4 star, archival data from the 6.5m Multiple Mirror Telescope (MMT, see Section 3.1).}}
\label{Figure3}
\end{figure*}

We reduce the data using a combination of custom-built {\sc idl} procedures and standard {\sc iraf} routines. In {\sc idl} we trim the spectrum, bias subtract and filter for cosmic rays, before we divide our data by a median averaged flatfield, which we first correct for dispersion effects using a pixel-integrated sensitivity function. We use {\sc iraf}'s {\sc apall} routine to perform optimal 1-D spectral extraction. Wavelength calibration is obtained using arc spectra from the standard CuNe-CuAr lamps and flux-calibration and telluric line removal is attempted with an early-type standard star closely matched to the target in airmass. The final spectrum is shown in Figure \ref{Figure3} as the black continuous line. In the upper panel of this figure, we indicate a few important molecular absorption bands, and several atomic lines. A best-fit model spectrum (solid red curve) and a best-matching observational template (solid green curve) are also overplotted on the lower panel of Figure \ref{Figure3}, which are discussed in further detail in Section 3.1. This spectrum is taken at 0.4 in binary phase, and although at this phase possible H$_\alpha$ emission of the secondary star may be blended with the primary we argue in Section 5.3, using medium resolution H$_\alpha$ spectral observations taken near quadrature, that such a contribution is likely very small and the shift of the H$_\alpha$ emission line is consistent with the primary star's radial velocity curve, indicating that the secondary is very dim at these wavelengths and may not be a very active star.

\subsubsection{Intermediate resolution spectroscopy}
We obtain intermediate resolution spectra with ISIS on the WHT, in order to measure the binary's radial velocity curve. We use the red arm with the R1200R grating (spectral resolution R$\sim$9300, 0.26 \AA /pixel) centred on 8500 \AA, giving a wavelength coverage of $\sim$8100-8900 \AA. This wavelength region (in particular the 8700-8850 \AA$ $ part) contains a number of relatively strong metallic absorption lines in M-dwarf atmospheres. Using this setup we obtain 5, 60 min (3$\times$20min) exposures on the nights of July 17 and July 18, 2010. Exposures of 90 min (3$\times$30 min) and 85 min (1$\times$1200+2$\times$1800) are also acquired with the same grating, but centred on the H-alpha emission line (6563 \AA), on the night of July 28, 2010. All observations are centered around the two quadrature points of the binary. Data reduction is performed using standard {\sc iraf} procedures for instrumental signature removal (the {\sc ccdproc} package), with bias subtraction, flat fielding and correction for instrumental response. We then calibrate the observed wavelengths using CuNe+CuAr arc lamp spectra taken after each set of exposures. We flux calibrate the data using observations of a photometric standard. Our spectra have a relatively low signal-to-noise ratio of (S/N)=15, meaning that because of the low luminosity of the secondary star in the optical, its spectral lines are not detected in this dataset. We use these observations to further constrain the radial velocity amplitude of the primary component as well as the orbital eccentricity in Section 3.
\subsection{GEMINI/GNIRS infrared spectroscopy}
We conduct observations with the GEMINI Near InfraRed Spectrograph (GNIRS) on the 8.1 m GEMINI-North Telescope, in queue-schedule mode in the H-band, to measure the secondary RVs. We move observations into the infrared, because the cooler companion is brighter at these wavelengths and the estimated binary luminosity ratio in the infared H-band is a factor $\sim$3 higher than in the optical red spectra. We opt to use an intermediate resolution setup, rather than a high resolution setup, to maximise the spectral throughput for the secondary star, which results in a compromise on the velocity resolution. We use the long slit (49``) and the long-red camera configuration, in combination with the 110.5 lines/mm grating and a slit-width of 0.3``, achieving a spectral resolving power of R$\sim$5900. Our set-up is centred on $\sim$1.555$\mu$m, and has a wavelength range of 1.49-1.61$\mu$m. The corresponding velocity resolution is $\sim$27.3 km/s/pixel. We observe our binary target using a standard $ABBA$ on-source ditther pattern, nodding along the slit. In total, we obtain 86, 240 sec. exposures on five seperate observing nights between March and July, 2011. Of these 86 exposures, 79 have sufficient signal-to-noise for radial velocity work. The observations are grouped as follows (corresponding to concurrent observations excluding low signal to noise data): 6$\times$240 s on the night of March 31, 2011, 12$\times$240 s on May 31, 12$\times$240 s on June 1, 12$\times$240 s on June 23 and finally 17$\times$240s and 21$\times$240s on the night of June 17. All observing runs are centered around the two quadrature points of the binary (phases $\phi$=0.25 and 0.75), where the relative RVs are expected to be the largest. 

We reduce the data using the GNIRS sub-package (version v1.11.1) of the Gemini {\sc iraf} package for spectral reduction and extraction, which is available online\footnote{Available at http://www.gemini.edu/sciops/instruments/gnirs/data-format-and-reduction}, and which is adapted by us for optimal reduction of the current data set. Before running the {\sc iraf} script we apply the {\sc python} script {\sc cleanir.py} to the raw data, to correct vertical striping (repetative every 8 columns), horizontal banding, and quadrant offsets, which represents a significant source of additional background noise in $\sim$70\% of the 2-D spectra, the magnitude of which also varies between different data frames. In {\sc iraf}, we first correct the data for read out noise, the detector offset (which is measured from a dark area of the data), and non-linear response using {\sc nsprepare}. Using {\sc nsreduce} we then remove the instrumental signature by flatfielding and dark subtraction, followed by sky-subtraction by forming $A-B$ and $B-A$ pairs. On each 
sky-subtracted 2-D image we then measure the dispersion variations using {\sc nssdist}, and rectify the images using {\sc nstransform}. With {\sc nscombine} each $A-B$ is then combined with the corresponding $B-A$ by shifting the positive spectra on to each other, based on the header information. Because the raw data show frequent spikes from radioactive particle hits, caused by decaying thorium on the lenses used in GNIRS, we median combine all sky-subtracted (A-B,B-A) pairs, which removes most of the hits. Our observations (especially the part taken at high airmass) also suffer from highly variable sky-lines from OH sky-glow, which causes line residuals even after sky-subtraction. Optimal 1-D spectral extraction is then performed using {\sc nsextract}, which is based on the standard {\sc iraf} package {\sc apall}. Any remaining particle hits are identified by eye from the 1-D spectrum and clipped before further 
analysis. A summary of the spectral observations is shown in Table 3.
\begin{table*}
   \begin{tabular*}{1.0\textwidth}
   {@{\extracolsep{\fill}}llllllll}
   \hline \hline
   Setup	&HJD	        &Slit(")	&$\lambda_{cen}$   &$t_{int}$	&Phase	&$\rm RV_1$	&$\rm RV_2$\\
   		&-2455000	&		&		   &	        &	&		&\\
   		&(days)		&		&(\AA)		   &(s)	        &	&(km/s)		&(km/s)\\
   \hline
   ISIS-R158R	&394.68899774	&1.0		&7250              &500		&0.400	&--	&--\\
   \hline
   ISIS-R1200R  &395.45466581       &1.2   &8495         &3$\times$1200	&0.713	&+58.5($\pm$1.6)	&--\\
   ISIS-R1200R  &395.58494467       &1.2     &8495       &3$\times$1200	&0.766	&+58.4($\pm$1.6)	&--\\
   ISIS-R1200R  &395.69615026       &1.2      &8495      &2$\times$1200	&0.812	&+57.2($\pm$1.8)	&--\\
   ISIS-R1200R  &396.56769165     &0.7      &8495      &3$\times$1200	&0.168	&+3.7($\pm$0.8)	&--\\
   ISIS-R1200R  &396.69054421      &1.0      &8495      &3$\times$1200	&0.219	&-0.1($\pm$1.0)	&--\\
   ISIS-R1200R    &406.41309437 &1.0   &6562      &1$\times$1200  &0.201	&+0.2($\pm$0.4)	&--\\
   		&	&	    &	&$\rm  +2\times1800$	&	&\\	
   ISIS-R1200R          &406.58030275 &1.0   &6562      &3$\times$1800  &0.269	&0.5($\pm$0.3)	&--\\
   \hline
   GNIRS	&713.9130117  &0.3	&15500	&12$\times$240 	&0.130	&+5.8($\pm$2.1)	&--\\
   GNIRS	&736.0121099  &0.3	&15500	&12$\times$240 	&0.185	&-1.9($\pm$2.7)	&--\\
   GNIRS	&711.3303809  &0.3	&15500	&12$\times$240 	&0.277	&+1.8($\pm$3.0)	&+135.5($\pm$3.0)\\
   GNIRS	&729.9526121  &0.3	&15500	&21$\times$240 	&0.680	&+56.5($\pm$4.0)	&-79.6($\pm$4.8)\\
   GNIRS	&730.0746367  &0.3	&15500	&17$\times$240 	&0.730	&+60.1($\pm$6.5)	&-80.1($\pm$4.2)\\
   GNIRS	&652.0859376  &0.3	&15500	&6$\times$240 	&0.820	&+51.5($\pm$3.0)	&-68.7($\pm$1.4)\\
   \hline
\end{tabular*}
\caption{\small{A summary of the spectral observations obtained with WHT ISIS and Gemini GNIRS. For the R1200R ISIS setup, the spectrograph is centered on 8495 \AA, and around the H$_{\alpha}$ emission line at central wavelength 6562 \AA. The columns $\rm RV_1$ and $\rm RV_2$ indicate the radial velocity derived from the primary and secondary line shifts respectively, uncorrected for the systemic velocity of the binary system, but converted to the heliocentric system, as in Figure \ref{Figure4}.}}
\label{proptable}
\end{table*}   

\section{Spectroscopic analysis}
\subsection{Analysis of the low resolution ISIS spectrum}
We determine the absorption indices in the red part of the optical spectrum of Sodium (Na$_{8189}$) and Titanium Oxide (TiO$_{7140}$) to constrain the luminosity class, and verify the main-sequence dwarf nature of WTS 19g-4-02069. We follow the procedure as outlined in Figure 11 of Slesnick, Carpenter \& Hillenbrand (2006), to distinguish low surface gravity giants from high gravity dwarfs. We find indices Na$_{8189}\sim$0.9 and TiO$_{7140}\sim$1.9, which are consistent with a main-sequence dwarf of spectral type $\sim$M3.5 (which is within the typical 1$\sigma$ uncertainties of the Reid, Hawley \& Gizis 1995 spectral type relation for TiO). We use the metallicity index $\zeta_{TiO/CaH}$ (described by Reid et al. 1995, Lepine, Rich \& Shara 2007 and Dhital et al. 2012), to find that the binary has solar metallicity within the uncertainties. Significant $\rm H_{\alpha}$ emission is seen (we estimate an equivalent width EW$_{H\alpha}$=-6 \AA) which is probably caused by the magnetic activity of the binary, related to the chromospheres of the stars. The strong presence of the Na I doublet, the absense of deep infrared Calcium triplet absorption (8498, 8542 and 8662 \AA), and lack of significant lithium absorption (6708 \AA), indicates that the primary is likely a mature M-dwarf, and not young and actively accreting nor a brown dwarf (e.g. Rebolo, Martin \& Magazzu 1992).

\begin{figure*}
\includegraphics[width=1.\textwidth]{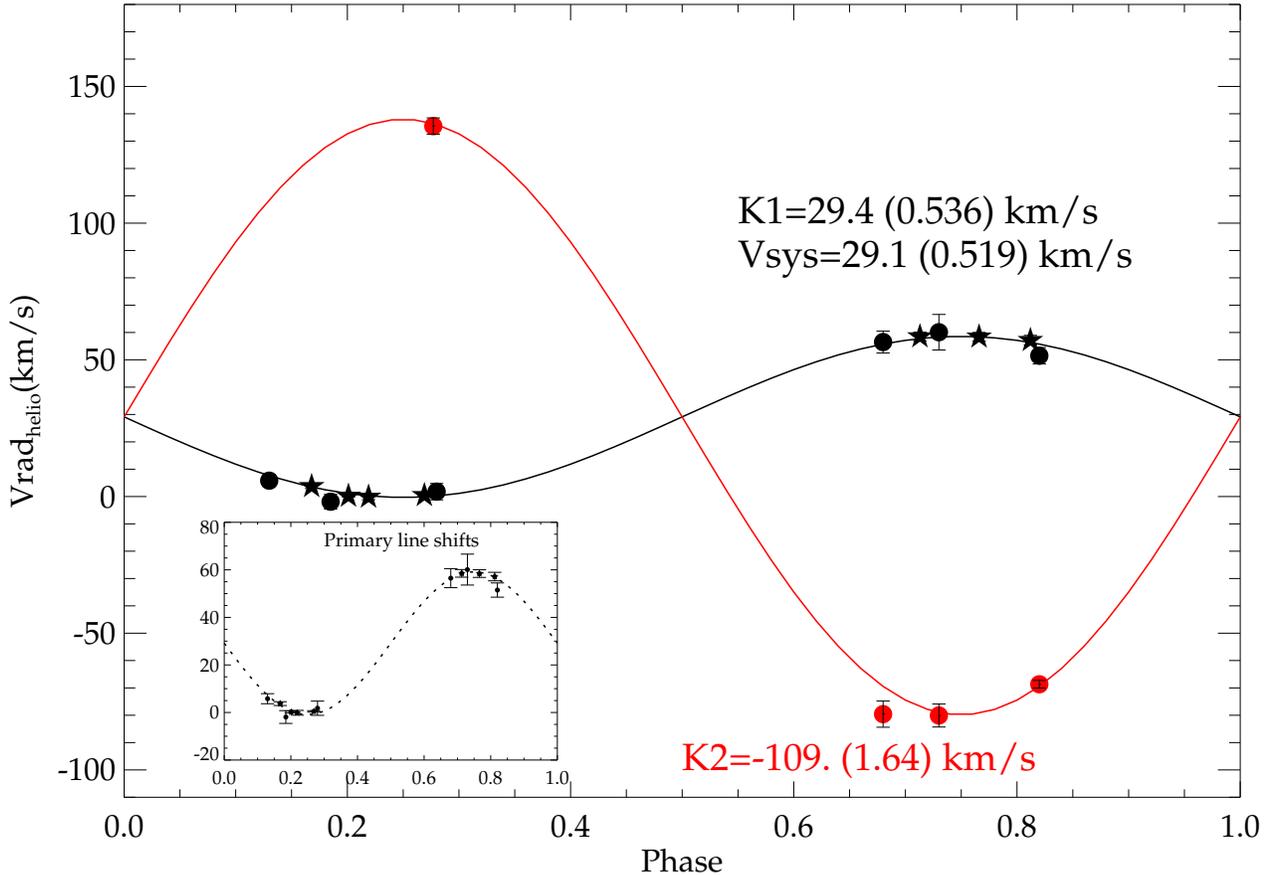}
\caption{\small{Radial velocity data for the primary (black filled symbols) and the secondary (red filled dots) stars of M-dwarf binary WTS 19g-4-02069 using GNIRS. Black filled stars are data from WHT ISIS, dots are from GEMINI GNIRS. The black and red solid curves are the best-fit sine functions. We obtain a binary systemic velocity of 29.1 ($\pm 0.5$)km/s. The data in the figure are corrected to the heliocentric system.}}
\label{Figure4}
\end{figure*}

We follow the procedure outlined in Nefs et al. (2012) and Birkby et al. (2012), to determine the system effective temperature, which can be used to derive individual component temperatures $T_{1,2}$ when combined with the light curve parameters, by $\chi^2$ fitting of a grid of NextGen atmosphere models (Allard et al. 1997) to the low resolution spectrum. This grid consists of models with $\Delta T_{eff}$=100K, and assumes constant $log(g)=5$ (typical for field M-dwarfs) and solar metallicity, spanning the $\sim$6000-9000 $\rm \AA$ region, which corresponds to the data-range least affected by instrumental effects. The model spectra are scaled to match the continuum of the observed spectrum. In the fitting procedure, we mask the strong telluric Oxygen bandhead around 7600 $\rm \AA$ and the $H_{\alpha}$ emission line. We use the formal errors as obtained from {\sc iraf} to derive the $\chi^2$, which we then optimise to determine the best-fitting model. To derive an error, we scale to $\chi^2_{red}$=1, yielding a final value of 140 K. Our best-fit model indicates $T_{eff}$=3200 K, which is consistent with a $\sim$M3-4 spectral type, following the $T_{eff}$, spectral type relation from Baraffe \& Chabrier (1996). Assuming that the primary star dominates the emission of the system in the optical, this spectral type roughly corresponds to that of the hotter component. The spectrum of GJ1001 (a $\sim$M4 single nearby M-dwarf\footnote{Available at http://spider.ipac.caltech.edu/staff/davy/ARCHIVE\\/index.shtml}), which was the best-fit observational template from a grid of M-dwarf spectra of various spectral subtypes observed with the Multiple Mirror Telescope (MMT), is overplotted on the lower panel of Figure \ref{Figure3} as a green continuous line. It is a better match to the ISIS data in the spectral regions around 6900 $\rm \AA$ than the NextGen model. 

We therefore identify WTS 19g-4-02069 as a genuine main sequence M-dwarf binary, with $\sim$M3.5 and $\sim$M5 components, and derive a best-fit effective binary temperature of 3200($\pm$140)K. Its stars have nearly solar metallicity and the system is significantly active with an EW$_{\rm H_\alpha}$=-6 \AA.

\subsection{Radial velocities}
\subsubsection{WHT/ISIS}
We use the {\sc iraf} routine {\sc fxcor} in conjunction with a grid of template synthetic stellar atmospheres of low-mass stars from the $MARCS$\footnote{Available at http://marcs.astro.uu.se/} spectral library (Gustafsson et al. 2008), degraded to match the resolution of the observed data, to obtain radial velocities from the R$\sim$9300 ISIS spectra through 1-D cross-correlation. In the cross-correlation procedure, we mask out the saturated near-infrared Ca II triplet lines at 8498, 8542 and 8662 $\rm \AA$. The spectral templates have a plane-parallel atmospheric geometry, an $T_{eff}$ range of 2800-4000 K (in steps of 100 K), solar metallicity, log($g$)=5 and 2.0km/s microturbulence. For the final RVs we use the template model that maximises the strenght of the cross-correlation, which is the cool $T_{eff}$=3200 K model. For the two $\rm H_{\alpha}$ observations we simply fit single Gaussians to the emission line. The data is listed in Table 3 and plotted in Figure \ref{Figure4} as black filled stars. Although radial velocities measured from H-alpha lines can be biased depending on where the H-alpha emission region is located, we find that our two radial velocity measurements are in good agreement with our red optical measurements and the line shape nearly Gaussian with no apparent major substructure. From the primary RVs we can already set reasonable upper and lower limits on the mass ratio of the system, because there is only a limited range of ($M_1,M_2$) that can yield the observed $K_1$. We find an upper limit, assuming $M_1=$0.08$M_{\odot}$ (corresponding to the hydrogen burning limit), of $q<$0.55, and a lower limit, assuming an M0-dwarf primary with $M_1=$0.65$M_{\odot}$, of $q>$0.25. This already indicates that WTS 19g-4-02069 is in the interesting regime of short period low-mass-ratio M-dwarf binaries.
\subsubsection{GEMINI/GNIRS}
\begin{figure}
\includegraphics[width=0.5\textwidth]{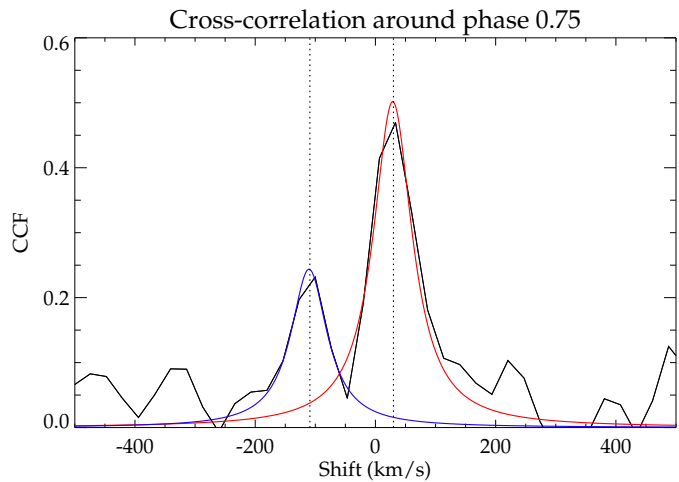}
\caption{\small{{The cross-correlation function for the summed spectral data for GNIRS Gemini around the 0.75 quadrature point using a template spectrum with $T_{eff}$=3200K, corrected for the Solar motion and the systemic velocity of the binary. The red and the blue solid curves indicate the best-fit single Lorentzian profiles for the primary and secondary lines.}}}
\label{Figure5}
\end{figure}

We use the spectral region 1.55-1.6$\mu$m, which is the least contaminated by telluric water vapor absorption, to extract RVs from the GNIRS data. With the IDL procedure $c\_crosscorrelate$ and a grid of MARCS model spectra we obtain a cross-correlation function (CCF) for each spectrum, which we subsequently fit by a Lorentzian. To improve the contrast of the CCF, we first normalise the spectrum by a second order polynomial. We evaluate for each combination of template spectra for the binary components the corresponding strength of the CCF. We find that the T=3200 K  model provides the highest signal for the primary of WTS 19g-4-02069, whereas for the secondary a cooler template of T=3000 K maximizes its cross-correlation signal. Given the large orbital velocities, there is relatively little blending of correlation peaks. We report the detection of a clear secondary component in the CCF around the $\phi=$0.75 quadrature point, shifted by $\sim$4.5 spectral pixels. We show the CCF of the total added spectral data around $\phi=$0.75 in Figure \ref{Figure5} (black curve). The blue and red solid curves indicate best-fitting single Lorentzian profiles to the cross-correlation peaks of the secondary and primary binary components respectively. The data around the 0.25 quadrature suffer from higher noise due to highly variable sky and detector pattern noise, making it difficult to resolve the secondary lines. Only the datapoint at phase 0.28 has sufficient quality for radial velocity work on the secondary.

We fit the RVs as function of binary orbital phase $\phi$ with a simple sine curve using {\sc idl}'s {\sc mpfitfun} function, assuming a circular orbit. We first fit the primary RVs from ISIS and GNIRS, allowing only $K_1$ and $\gamma$ (the systemic velocity of the binary) to vary, fixing the phase using the well-determined orbital period from the light curve. For the secondary star we fix $\gamma$ to the value derived from the primary and fit for $K_2$ (note that we obtain consistent results when we treat $\gamma$ as a free parameter). To derive the RV errors we scale the errors from the Lorentzian/Gaussian CCF fit, such that the reduced $\chi^2$ of the best-fitting RV model is unity. We show our results in Figure \ref{Figure4}, where the black solid curve indicates the best-fitting sine function to the primary RVs, and the red solid curve the fit to the secondary RVs. We find radial velocity amplitudes of $K_1$=29.4($\pm$0.5) km/s and $K_2$=109.0($\pm$1.6) km/s, for a systemic velocity $\gamma$=29.1($\pm$0.5) km/s, indicating a binary mass-ratio $q$=0.27($\pm$0.02). 
\begin{table*}
   \begin{tabular*}{1.0\textwidth}
   {@{\extracolsep{\fill}}llll}
   \hline \hline
   Parameter	&i-band	&J-band	&J-band\\
   		&	&	&(spot-corrected)\\
   \hline
   {\bf{light curve analysis}}	&	&\\
   P (days)	&2.44178($\pm0.00003$)	&	&\\
   $T_0$ (MJD)	&2454319.83270($\pm0.00002$) &	&\\
   $(R_1+R_2)/a$	&0.103($\pm$0.002) &0.104($\pm$0.002)	&0.098($\pm$0.004)\\
   $R_2/R_1$	&0.338($\pm$0.003)         &0.343($\pm$0.003)	&0.339($\pm$0.005)\\
   $J$	&0.40($\pm$0.01)	               &0.65($\pm$0.01)	&0.64($\pm$0.02)\\
   $i(^{\circ})$     &87.7($\pm$0.1) &87.7($\pm$0.2)	&88.2($\pm$0.3)\\
   $R_1/a$	&0.077($\pm$0.001)	         &0.077($\pm$0.001)	&0.073($\pm$0.003)\\
   $R_2/a$	&0.026($\pm$0.001)	                 &0.026($\pm$0.001)	&0.025($\pm$0.003)\\
   $L_2/L_1$	&0.042($\pm$0.004)	                 &0.074($\pm$0.004)	&0.072($\pm$0.007)\\
   $T_2/T_1$	&0.779($\pm$0.008)	                 &0.892($\pm$0.007)	&0.89($\pm$0.01)\\
   $\chi^2_{red}$	&1.093		                 &4.057	                &1.004\\
   {\bf{Estimated temperatures}}	&	&	&\\
   $T_{eff} (K)$	&3200($\pm$140)	&	&\\
   $T_1 (K)$	&3300($\pm$140)		&	&\\
   $T_2 (K)$	&2950($\pm$140)		&	&\\
   \hline
   {\bf{Radial velocity analysis}}	&	&	&\\
   $K_1$(km/s)	&29.4($\pm$0.5)	&	&\\
   $K_2$(km/s)	&109.0($\pm$1.6)	&	&\\
   $\gamma$(km/s)	&29.1($\pm$0.5)	&	&\\
   $q$		&0.27($\pm$0.02)		&	&\\
   a($R_{\odot}$)	&6.7($\pm$0.2)	&	&\\
   \hline
   {\bf{Derived masses and radii}}	&	&	&\\
   $M_1$ ($M_{\odot}$)	&{\bf{0.53($\pm$0.02)}}	&0.53($\pm$0.02)	&0.53($\pm$0.02)\\
   $R_1$ ($R_{\odot}$)	&{\bf{0.51($\pm$0.01)}}	&0.52($\pm$0.01)	&0.49($\pm$0.02)\\
   $M_2$ ($M_{\odot}$)	&{\bf{0.143($\pm$0.006)}}	&0.143($\pm$0.006)	&0.143($\pm$0.006)\\
   $R_2$ ($R_{\odot}$)	&{\bf{0.174($\pm$0.006)}}	&0.177($\pm$0.006)	&0.167($\pm$0.009)\\
   \hline
   \end{tabular*}
\caption{Best-fit parameters and derived quantities for the M-dwarf binary system WTS 19g-4-02069. The boldface fonts indicate the adopted masses and radii derived from the INT i-band light curve parameters (see the discussion in Section 4.3). }
\label{proptable}
\end{table*}

\section{Light curve analysis}
\subsection{J-band photometry baseline variation analysis}
\begin{figure*}
\begin{center}$
\begin{array}{ll}
\includegraphics[width=0.5\textwidth]{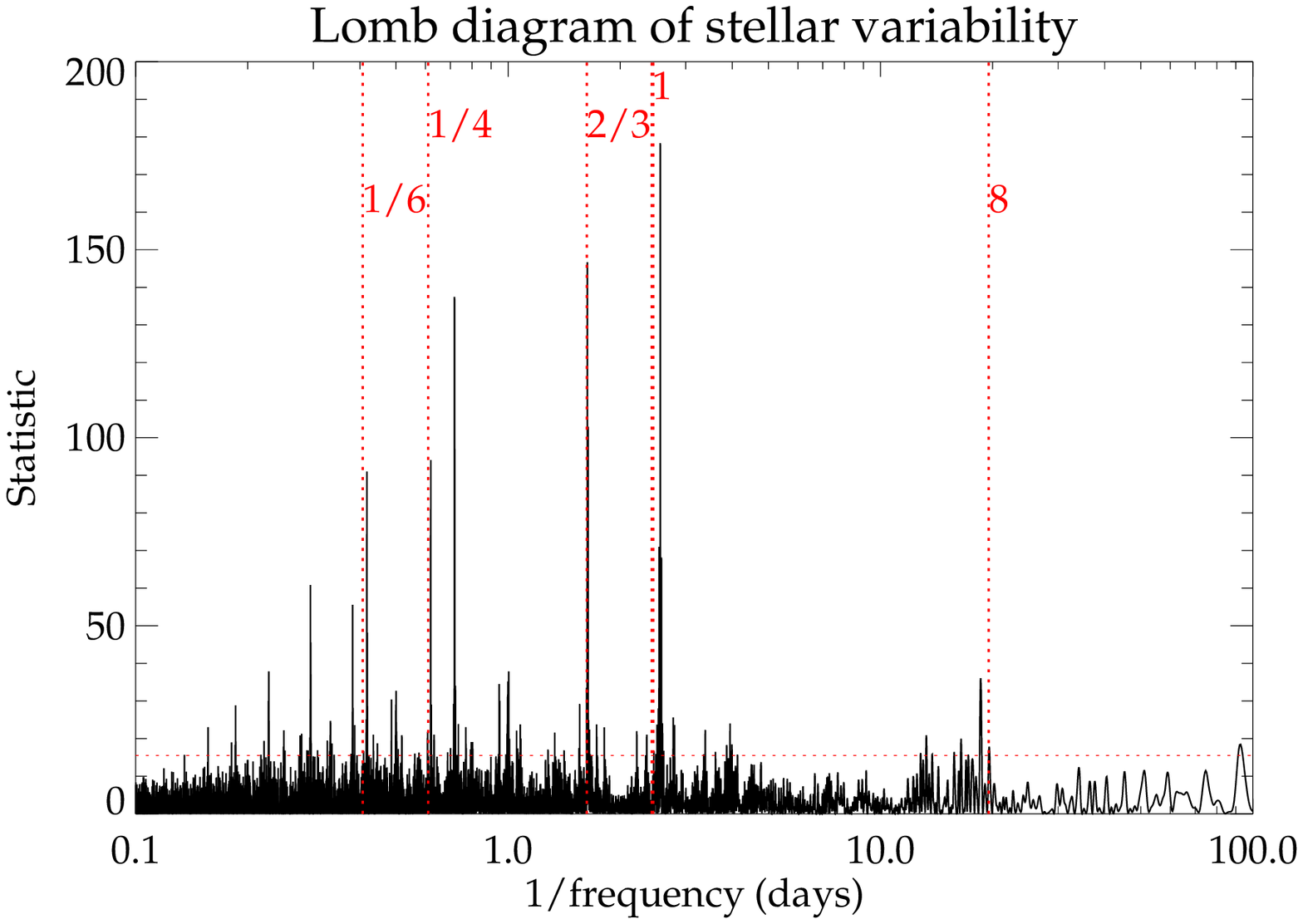} &
\includegraphics[width=0.5\textwidth]{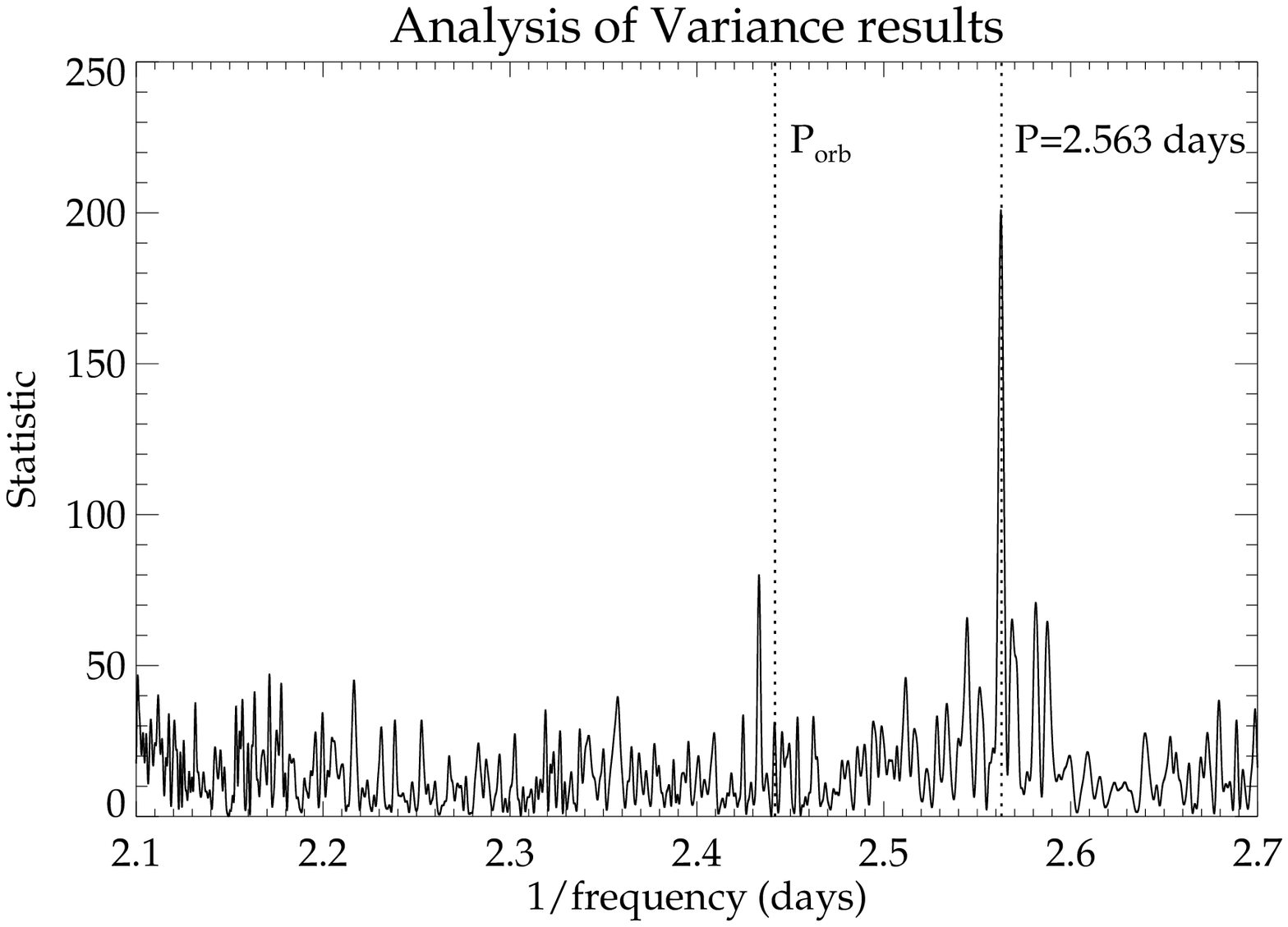} \\
\end{array}$
\caption{\small{Left panel: Lomb-Scargle diagram of the dominant frequencies in the WFCAM J-band light curve after removing the eclipses. A significant peak is at $\sim$2.563 days. We indicate several period aliases with the red dashed lines. The red horizontal line indicates the 99\% confidence level for peak rejection. Right panel: Analysis of Variance results for binary WTS 19g-4-02069. We show the frequency spectrum between 2.1 and 2.7 d and note that we confirm a strong signal at the period determined by the Lomb-Scargle method, and no significant signal at the orbital period $P_{orb}\sim$2.44 d. The second most significant peak is at $\sim$2.432 d, but with low significance. }}
\label{Figure6}
\end{center}
\end{figure*}

There is significant out-of-eclipse scatter in the J-band light curve ($\sim$2\% peak-to-peak), which is not in phase with the binary orbit. To investigate the possible periodicity of this variation, we clip the primary and secondary eclipses from the light curve, and perform a frequency analysis using the {\sc idl} implementation {\sc fasper.pro} of the Lomb-Scargle periodogram (Lomb 1976; Scargle 1982; Press \& Rybicki 1989). The left panel of Figure \ref{Figure6} shows this frequency spectrum for binary WTS 19g-4-02069. Using the {\sc idl} routine, we determine a false-alarm probability (FAP) to filter out peaks that are likely caused by spurious detections on light curve systematics (horizontal dashed line). Significant power is apparent around $\sim$2.44 d binary period, yet the actual peak is $\sim$0.14 d away at $\sim$2.56 d. We indicate various integer and half-integer aliases of the binary orbital period in Figure \ref{Figure6} using vertical dotted lines. From previous binary studies, similar deviations are seen in systems that are either young and not fully synchronised (e.g. the pre-main sequence eclipsing binary Paranengo 1802; Cargile et al. 2008) or not fully circularised due to their relatively long orbital period (e.g. Irwin et al. 2011). We discuss possible causes for this apparent discrepancy in Section 5.

When folded onto the non-synchronous 2.56 d period determined from the Lomb-Scargle analysis, we find a clear nearly-sinusoidal modulation of the data, which we attribute to star spots on a rotating stellar surface. We also perform an independent check of these results using the {\sc idl} implementation {\sc epfold} of the Analysis of Variance (AoV;Schwarzenberg-Czerny 1989) algorithm and plot the results in the right panel of Figure \ref{Figure6}. We confirm a best-fit period which is very close to the $\sim$2.563 d period suggested by the Lomb-Scargle algorithm. No significant signal is obtained at the orbital period. The second highest peak in the right hand panel is at $\sim$2.432 d, but at relatively low significance. In Figure \ref{Figure7} we show the clipped light curve folded on the best-fit periodicity for spot modulation. We attempt to correct the J-band light curve using a single sine with 8.1 mmag amplitude, which appears to minimise the out-of-eclipse rms. In the upper left panel of Figure \ref{Figure1} we show the phase-folded but uncorrected light curve (black filled dots), while in the upper right panel panel we show the J-band data, corrected for the $\sim$8.1mmag rotation signal. We do not claim that this method is the best method for removing rotational signal from a light curve, because if spots are occulted during the eclipse the light variation may strongly deviate from the sinusoid. However, we find that with our correction the out-of-eclipse rms decreases from $\sim$10.6 to 9.1 mmag ($\sim$15\% reduction), and in eclipse it decreases from $\sim$11.2 to 8.9 mmag ($\sim$20\% reduction). This means that the rms in and out of eclipse are approximately the same after the correction (in fact the difference between the rms values in and out of eclipse decreases from $\sim$5\% to $\sim$2\% after correction). Zoom-ins around the eclipses of the corrected and uncorrected data are shown in Figure \ref{Figure1}.

\subsection{{\sc jktebop} light curve fits}
\subsubsection{J-band photometry}
We use the binary light curve modelling program {\sc jktebop}\footnote{Available at http://www.astro.keele.ac.uk/~jkt/} (Southworth et al. 2004), which is based on the Eclipsing Binaries Orbit Program ({\sc ebop}; Popper \&Etzel 1981; Etzel 1981) for the fitting of the uncorrected and corrected J-band data. For binary WTS 19g-4-02069, with orbital period 2.44 d and mass ratio 0.27, Equation 6 of Morris (1985) predicts ellipsoidal light variations of just $\sim$0.4 mmag in the J-band, indicating that the binary stars are likely only very slightly deformed by mutual tidal interactions. This, together with the low derived values of the stellar oblateness in our subsequent fitting with {\sc jktebop}, justifies the application of this model (which is only suitable for detached systems) to the WTS 19g-4-02069 binary system (see also Popper \& Etzel 1981). 

For the light curve modelling we allow the following six parameters to vary: i) the binary orbital period $P_0$, ii) the mid-eclipse epoch $T_0$ of the primary eclipse, iii) the sum of the stellar radii in units of the binary semi-major axis, $(R_1+R_2)/a$, iv) the ratio of the radii $k=R_2/R_1$, v) the orbital inclination $i$ and vi) the ratio of central surface brightness $J=J_2/J_1$. We use as input the initial estimate of the binary ephemeris obtained from the automated {{\sc occfit}} BLS algorithm. 

We keep the mass-ratio of the system, $q$, fixed at the value determined by the spectroscopic analysis. We do not fit the reflection coefficients, but calculate them from the system geometry. We also assume a gravity darkening coefficient, which is fixed at a value typical for stars with convective envelopes ($\beta$=0.32; Lucy 1967). We find that treating additional third light as an extra fitting parameter does not significantly improve the quality of our fit, so we fix it to zero. We adopt linear limb-darkening coefficients from Claret et al. (2000) in the J-band (see Table \ref{limbdarkening}), which are calculated from PHOENIX model atmospheres (Allard et al. 1997), for a surface gravity $log(g)=5$, solar metallicity and 2 km/s micro-turbulence and stellar effective temperatures $T_{1,2}$, such as derived in Section 4.2.2. We do not fit for the limb-darkening coefficients, because the S/N is too poor, and keep $T_{1,2}$ (see Section 4.2) fixed. The orbital eccentricity $e$ and the argument of periapse $\omega$ are also kept fixed to values consistent with a circular orbit, because our initial runs indicate that the data are firmly consistent with such an orbit ($|e*cos(\omega)|<0.000079$). This is expected from tidal dissipation theory given the relatively short circularisation timescale $\sim$200 Myr (Zahn 1977). 
\begin{table}
   \begin{tabular*}{0.5\textwidth}
   {@{\extracolsep{\fill}}llll}
   \hline
           &		&$i$-band	&J-band\\
   \hline
   Linear  &Primary	&0.69		&0.41\\
   	   &Secondary	&0.78		&0.46\\
   Quadr.  &Primary	&[0.22;0.58]	&\\
   	   &Secondary	&[0.37;0.49]	&\\			 
  \hline
   \end{tabular*}
\caption{Limb-darkening coefficients used as input to the EBOP models for the WFCAM J-band and the INT i-band light curves. For the optical red data we indicate both linear and quadratic coefficients.}
\label{limbdarkening}
\end{table}

To assess the 1$\sigma$ parameter uncertainties we use the Monte Carlo routine from {\sc jktebop} (Southworth et al. 2005). In this procedure, Gaussian random noise is repeatedly (10000 times) added to the model light curve before a new fit is made to the data, which yields a distribution for each parameter. With {\sc jktebop} we also perform a prayer-bead error analysis, which can be useful in the presence of correlated noise (Southworth 2008), and find that parameter values derived from both methods are consistent within the 1$\sigma$ uncertainties, and that the uncertainties are not significantly different from the prayer-bead analysis. We therefore adopt the MCMC method from hereon. Numerical results of the light curve fitting are given in Table 4. 

The derived parameters for the uncorrected and the spot-corrected data are inconsistent within the quoted 1$\sigma$ uncertainties, $(R_1+R_2)/a$, $R_2/R_1$, and $J$ are smaller after correction, whereas $i$ is higher. In Figure \ref{Figure1} it is apparent that the correction removes outlying data points in eclipse, which bias the measured depths of the eclipses and the duration of ingress/egress. Note that the errors of the corrected light curve data points are scaled to obtain $X^2_{red}$=1. The derived ratio of radii ($R_2/R_1$=0.339$\pm$0.005) in the data indicates that the companion is significantly smaller than the primary. An impact parameter $b=(a/R_1)*cos(i)$=0.43 suggests that the eclipses are full and the system is non-grazing, which lifts the degeneracy between $R_2/R_1$ and $i$. The ratio of secondary to primary luminosity is $L_2/L_1$=0.072($\pm$0.007), which indicates that only $\sim$7\% of the system light in the infrared is due to the secondary star. From $L_2/L_1$ and $R_2/R_1$ we derive, using Stefan-Bolzmann's law, a wavelength specific temperature ratio $T_2/T_1$=0.89($\pm$0.01). Assuming that all of the light in the system comes from the primary, and the stars radiate as blackbodies, this indicates $T_2\sim$2850K for the secondary, which would be consistent with a M5 type star according to the $T_{eff}$-spectral type relation presented in Stephens et al. (2009), M5 according to the 1 Gyr model from Baraffe \& Chabrier (1996), and M6 according to Reyle et al. (2011). The mass and radius of the secondary, such as derived in Section 4.3, indicate a surface gravity log(g)$\sim$5.1 which is consistent with a main-sequence M-dwarf.
\begin{figure}
\includegraphics[width=0.5\textwidth]{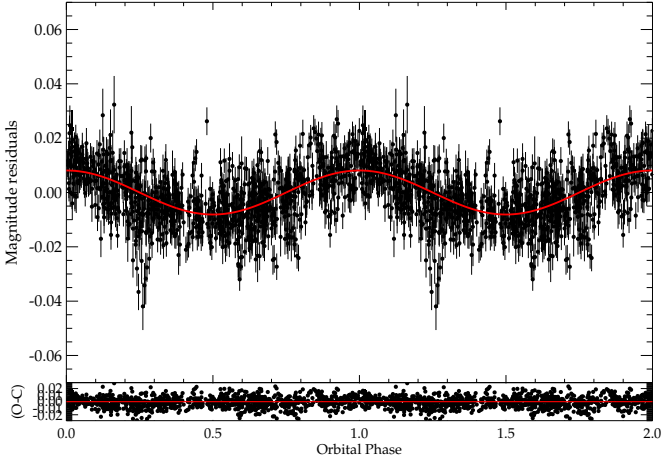}
\caption{\small{The clipped WFCAM J-band light curve folded on the $\sim$2.563 day frequency peak. A significant rotational modulation can be seen. The solid red curve is the best fit sinusoid to the data with amplitude $\sim$8.1mmag.}}
\label{Figure7}
\end{figure}

\subsubsection{i'-band photometry}
The lack of out-of-eclipse baseline for the optical data means that we can not accurately determine the amplitude of any spot modulation, so we opt to fit the light curve without making any spot corrections. The INT eclipses have a flat bottom, which confirms that the secondary star is fully superimposed on the primary during eclipse. To model with {\sc jktebop}, we adopt the linear coefficients from Claret et al. (2004) for the Sloan i'-band. The derived parameters $(R_1+R_2)/a$ and $R_2/R_1$ are consistent with the spot-corrected infrared results, within the quoted 1$\sigma$ uncertainties of the J-band data. The orbital inclination $i$ is slightly lower in the optical, but only by $\sim$1.4$\sigma$. The optical data reveal a surface brightness ratio $J$=0.40($\pm$0.01) and a 
wavelength specific luminosity ratio $L_2/L_1$=0.042($\pm$0.004). This result shows that the secondary is significantly dimmer at optical wavelengths. Also, a ratio $T_2/T_1$=0.779($\pm$0.008) is derived in the optical ($\sim$11\% lower than in the infrared). To derive component temperatures $T_{1,2}$ we use MARCS model spectra, convolved to the i' and J bandpasses, to derive model surface brightness ratios $J_{\lambda}$, which are compared to the optical and infrared observations. We reproduce the observations, within the estimated 1$\sigma$ uncertainties, for $T_1$=3300 K and $T_2$=2950 K , with an estimated uncertainty of 140 K. We adopt these values of $T_1$ and $T_2$ in our subsequent discussion.

\subsection{Stellar Masses and Radii}
We derive the component masses and radii from the combined RV analysis (incorporating both the ISIS and the GNIRS measurements), the i-band light curve fitting parameters, and Keplers law according to the following two equations (see e.g. Hilditch 2001):
\begin{equation}
M_1=\frac{K_2 P(K_1+K_2)^2}{2\pi Gsin(i)^3}
\end{equation}
\begin{equation}
R_1=x_1a=x_1\left[\frac{P^2GM_1(1+K_1/K_2)}{4\pi^2}\right]^{1/3},
\end{equation}
where $x_1$ denotes the best-fit scaled binary orbital separation, $R_1/a$, from {\sc jktebop}. We propagate the errors from the light curve and radial velocity analysis. The main motivation for using the i-band results is that for the infrared data, many spot cycles are folded into the light curve, and although our single sine correction removes part of the scatter, tracing the stellar activity cycle over such a long observational baseline ($\sim$5 years) is difficult because the spot configuration may have evolved significantly. Moreover, the J-band data has significant gaps in observing dates and generally only few observations per night, making it difficult to accurately model the precise behaviour of the spots. Because the optical data for the primary and the secondary event are obtained within 6 d of observations, they likely trace the same spot configuration. Moreover, both the photometric quality and the number of in-transit datapoints of the optical data can rival the J-band photometry. One problem in the current work is the limited coverage of the parts out of eclipse. Work by Goulding et al. (2012) shows that the light curve amplitude of a spot in the J-band is generally 55\% of that in the I-band, indicating that $\sim$3\% peak-to-peak variations would be expected for INT observations of the full binary orbit. This could introduce an additional error to our light curve fitting results. Future work should adress this issue by obtaining multiband photometric observations of concurrent binary eclipses and sufficient baseline, to catch a single spot cycle. With $x_1$=0.077, $K_1$=29.4km/s, and $K_2$=109.0 km/s, we find radii of $R_1$=0.51($\pm$0.01)$R_{\odot}$, $R_2$=0.174($\pm$0.006)$R_{\odot}$, and masses $M_1$=0.53($\pm$0.02)$M_{\odot}$, $M_2$=0.143($\pm$0.006)$M_{\odot}$. We use equations similar to Equations 1 and 2 to derive the mass and radius of the secondary. This translates to 2.1-3.2\% errors on the radii and $\sim$4\% errors on the masses, although we caution that these errors do not include possible uncertainties from star spots. Note that these masses and radii are consistent with main-sequence model predictions (see Section 5.2), rather than pre-main sequence, providing further support for the mature nature of the system. 

\section{Discussion}
In this paper we have presented the discovery of a highly unequal-mass eclipsing M-dwarf stellar binary, using the high-precision infrared light curves of the WFCAM Transit Survey, and follow-up characterisation with optical photometry and optical and IR spectroscopy on 2.5-8 m class telescopes. With two components straddling the fully convective boundary, and with shared ages and metallicities, our binary provides a rare and more stringent comparison to structure model predictions of fundamental M-dwarf properties over a wide span of stellar masses. The cool M5V secondary of the binary is in an important mass-regime for studies of exoplanets down to the Earth-size regime. In this section we will discuss our binary in the context of current theories for low-mass binary formation, which predict such close unequal systems to be rare and in the context of low-mass stellar structure models. 
\subsection{The mass-ratio distribution}
The distributions of binary orbital separation and mass-ratio, as function of primary mass, provide important constraints on star formation simulations (e.g. Burgasser et al. 2007, Bate et al. 2012, Clarke 2012). These simulations suggest that accreting gas with high angular momentum and dynamical interactions tend to drive up the mass-ratio of close binary systems towards unity. Dynamical interactions are frequent in the binary birth environment, and interactions with more massive stars will generally bias binary primaries towards higher masses. This indicates that it is unlikely that close, unequal binary systems with low-mass primaries can be maintained for very long. For example, a 0.5+0.1$M_{\odot}$ M-dwarf binary that is formed in a cluster with stellar density $n_*\sim$2000pc$^{-3}$ has a life-expectancy against disruption from solar type stars of order $\sim$10 Myr (Goodwin \& Withworth 2007). Furthermore, low-mass binary systems and unequal-mass systems are more easily perturbed due to their lower binding energy. 

\begin{figure*}
\begin{center}$
\begin{array}{ll}
\includegraphics[width=0.5\textwidth]{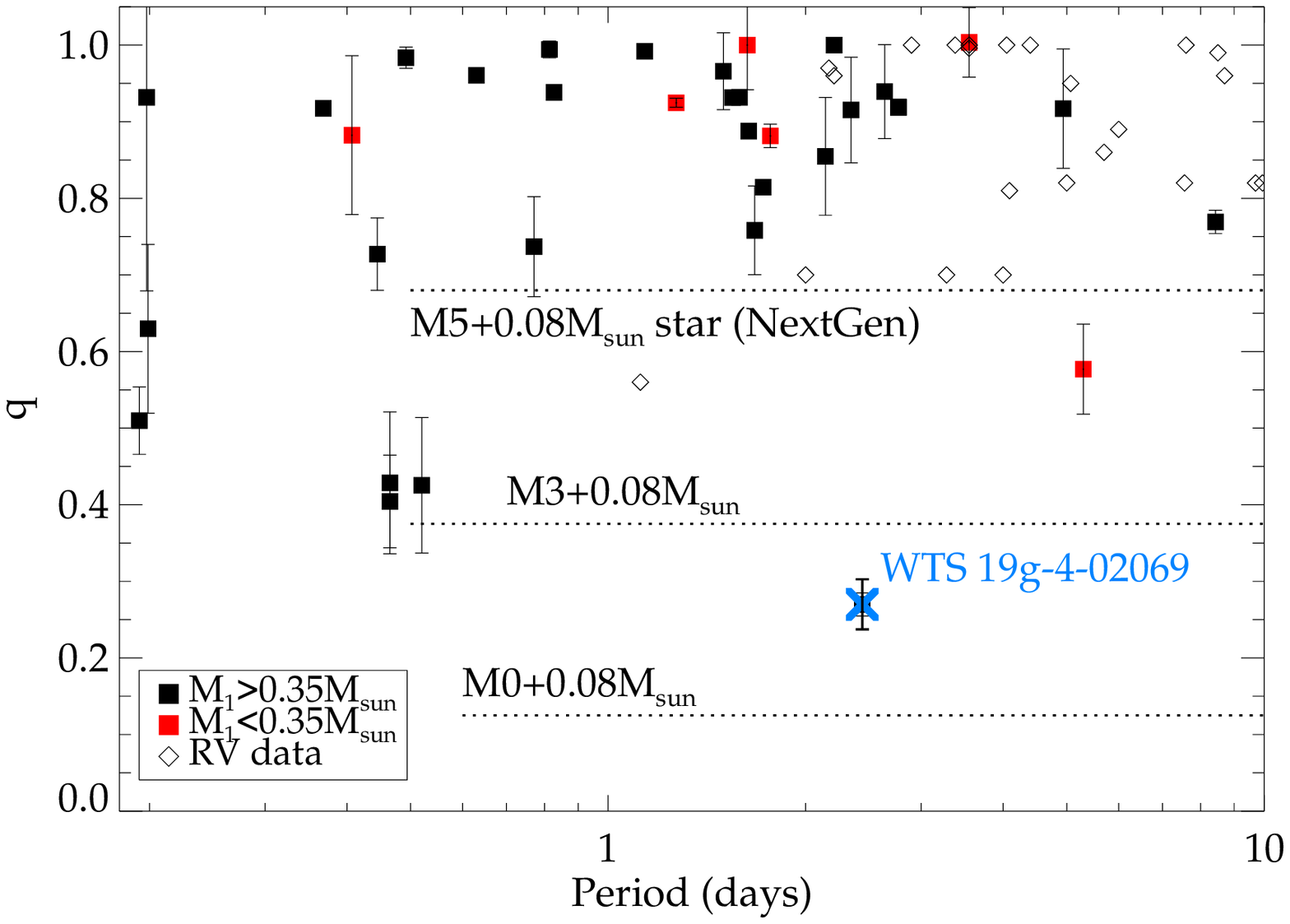}&
\includegraphics[width=0.5\textwidth]{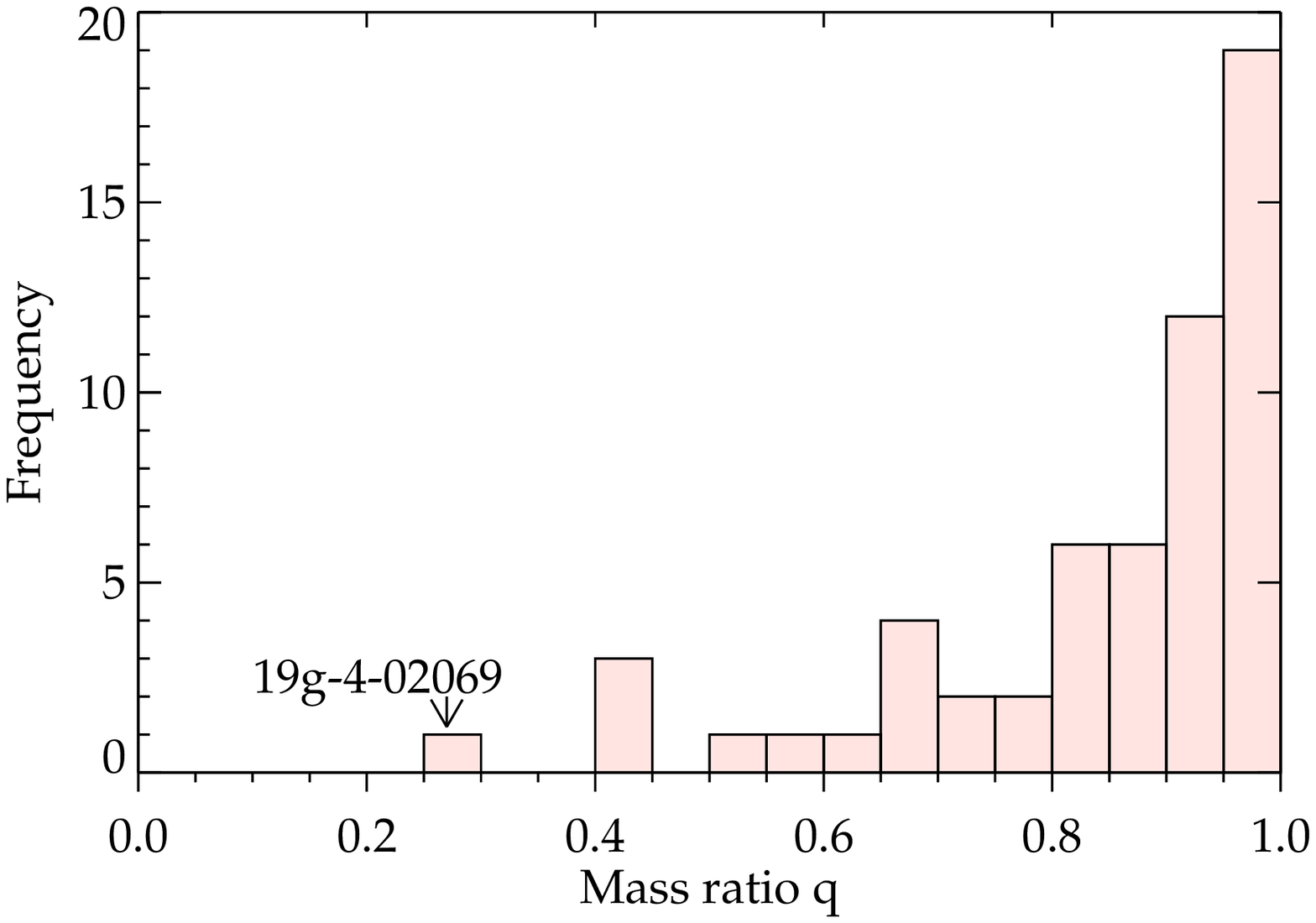} \\
\end{array}$
\end{center}
\caption{\small{The mass-ratio distribution of short period ($P<$10 d) M-dwarfs reported in literature. Left panel: binary orbital period (in d) versus mass-ratio. Black filled squares indicate eclipsing M-dwarf binaries with $M_1>0.35M_{\odot}$, red squares indicate eclipsing binaries with $M_1<0.35M_{\odot}$, whereas open diamonds show M-dwarf binaries reported in radial velocity surveys. The large blue cross is our measured mass-ratio for M-dwarf binary WTS 19g-4-02069. The three black dashed lines show the respective mass-ratios for M-dwarf binary systems with a secondary at the hydrogen burning limit, assuming primaries of 0.6$M_{\odot}$ (M0 spectral type; Baraffe \& Chabrier 1996), 0.2$M_{\odot}$ (M3), and 0.11$M_{\odot}$ (M5). These lines indicate the lower limits in $q$ to which M-dwarf primaries with stellar secondaries are confined. Right panel: histogram of mass-ratio.}}
\label{Figure8}
\end{figure*}

In Figure \ref{Figure8} we show the mass-ratio distribution of close binary systems ($P_{orb}<$10 d) with M-dwarf primaries. This figure is compiled from Table 6, which holds the currently available sample of M-dwarfs discovered as eclipsing and non-eclipsing, double-lined, spectroscopic binaries. This is an updated version of Figure 9 in Wisniewski et al. (2012), who include only a few M-dwarf systems in the eclipsing binary period range. The left panel of our Figure \ref{Figure8} shows mass-ratio as function of binary orbital period for 55 sources, of which 20 are spectroscopic binaries, whereas the right panel shows a histogram of the mass-ratio distribution in bins of 0.05. We separate the M-dwarfs depending on whether their primaries are more or less massive than the fully convective boundary at $\sim$0.35$M_{\odot}$ (black and red filled squares). With a mass-ratio of 0.27, it is clear that WTS 19g-4-02069 occupies an interesting position in these diagrams, because over 80\% of the stellar binaries have q$>$0.8. There may be an observational bias towards more equal-mass M-dwarf binaries because of the steep relation of mass and luminosity for M-dwarfs, causing the spectral lines of low-luminosity companions to remain unresolved with optical spectroscopy. Also, for lower q systems, lower radial velocity shifts of the primary lines are expected, making them more difficult to detect a priori, especially for binaries with late type M-dwarf primaries, which are intrinsically faint. In order to better understand this bias it is important to move future spectroscopic and photometric observations into the infrared, because of the improved binary luminosity ratio there. Also any single-lined short-period eclipsing systems should be reported, because for such systems an upper limit can be obtained on the mass-ratio given that the orbital inclination can be directly constrained from the light curve. Yet, note that in the Birkby et al. (2012) candidate MEB list, more particular the sample with best-fit binary temperature less than 4000 K, we do not identify any object with a ratio of secondary to primary eclipse depth significantly lower than that of binary WTS 19g-4-02069, even though we have the photometric precision to detect such low-mass M-dwarf companions through their eclipses. This may indicate that such systems are in fact intrinsically rare. As we will discuss in the next paragraph such considerations are important constraints on the physics of low-mass binary formation and gas accretion mechanisms.

Three binaries are observed in the range q=[0.4-0.5], but none at lower q. Of these three, two are very short-period ($\sim$0.4 d) and young, 20 Myr T-Tauri stars (NSVS-06507557; Cakirh \& Ibanoglu 2010) and 150 Myr young cluster members (2MASSJ04463285; Hebb et al. 2006). Theory shows that significant dynamical processing can occur prior to the main-sequence, indicating that low mass-ratio systems should be more abundant while young. For example, a significant difference in the binary fraction between young clusters and field solar type stars has been observed (e.g. Duchene et al. 2007). 

A possible explanation for the existence of WTS 19g-4-02069 is that the physics of gas accretion onto (close) binary systems is different than suggested by smoothed particle hydrodynamic (SPH) and ballistic particle simulations (from e.g. Artymowicz 1983; Bate \& Bonnell 1997; Bate 2002). Two-dimensional warm grid-based simulations from Ochi et al. (2005) and de Val-Borro et al. (2011) confirm that gas preferentially enters the secondary Roche lobe, but flows around the secondary and is then channeled onto the primary star, which grows a more massive accretion disk, which means that q can decrease during the accretion phase. Very recent numerical work from Zhao \& Li (2012) suggests that adding magnetic fields to binary formation simulations can have a similar effect. Such a magnetic field could apply a brake on the material that flows onto the binary, decreasing its angular momentum and significantly shrinking the protobinary separation, meaning that q could be low for short binary orbits. If these simulations are correct, low-q pairs are expected to be abundant around a wide range of primary masses on the main-sequence, which would be consistent with our observations of WTS 19g-4-02069. However, disrupting third-body interactions can still remove low-q components from binaries. Another explanation for the low mass-ratio of WTS 19g-4-02069 could be that the binary was isolated from the birth environment early on, for example through ejection from the cloud due to binary-single star or binary-binary interactions, or because the natal cluster became unbound due to rapid gas removal.   
\subsection{The mass-radius relation for M-dwarfs}
Highly unequal-mass M-dwarf binaries provide important test cases of low-mass stellar evolution theory because they cover a large range of M-dwarf masses, which encompass significant changes in stellar atmospheric structure. Furthermore, since the binary components have the same age and metallicity, stellar evolution models can be tested using two less free parameters. Also, Earth-like planets in the habitable zones around late-type M-dwarfs are prime targets for new transit surveys and Extremely Large Telescopes (ELTs), and require accurate calibration of their small host stars. For example, for the Neptune-sized planet orbiting the M5V host star GJ1214b (Charbonneau et al. 2009, de Mooij et al. 2012), a 15\% uncertainty in stellar radius could translate to the difference between an ocean planet and a gaseous H- or He-type atmosphere.

Theory predicts that stars in the fully convective mass regime ($<$0.35$M_{\odot}$) respond differently to rapid rotation and strong magnetic fields (which can reduce convective efficiency) than their partially-convective cousins (e.g. Chabrier et al. 2007). Most importantly, because in fully convective atmospheres heat flow is nearly adiabatic, model stellar radii and temperatures are expected to better match the observations. It is argued that strong magnetic fields can shift the fully convective mass boundary to masses as low as 0.1$M_{\odot}$ (Mullan \& MacDonald 2001). Such predictions can be tested with a sample of mass-radius-temperature measurements. However, in the stellar-mass regime $<$0.2$M_{\odot}$ there is a lack of model-independent data.

\begin{figure*}
\includegraphics[width=1.\textwidth]{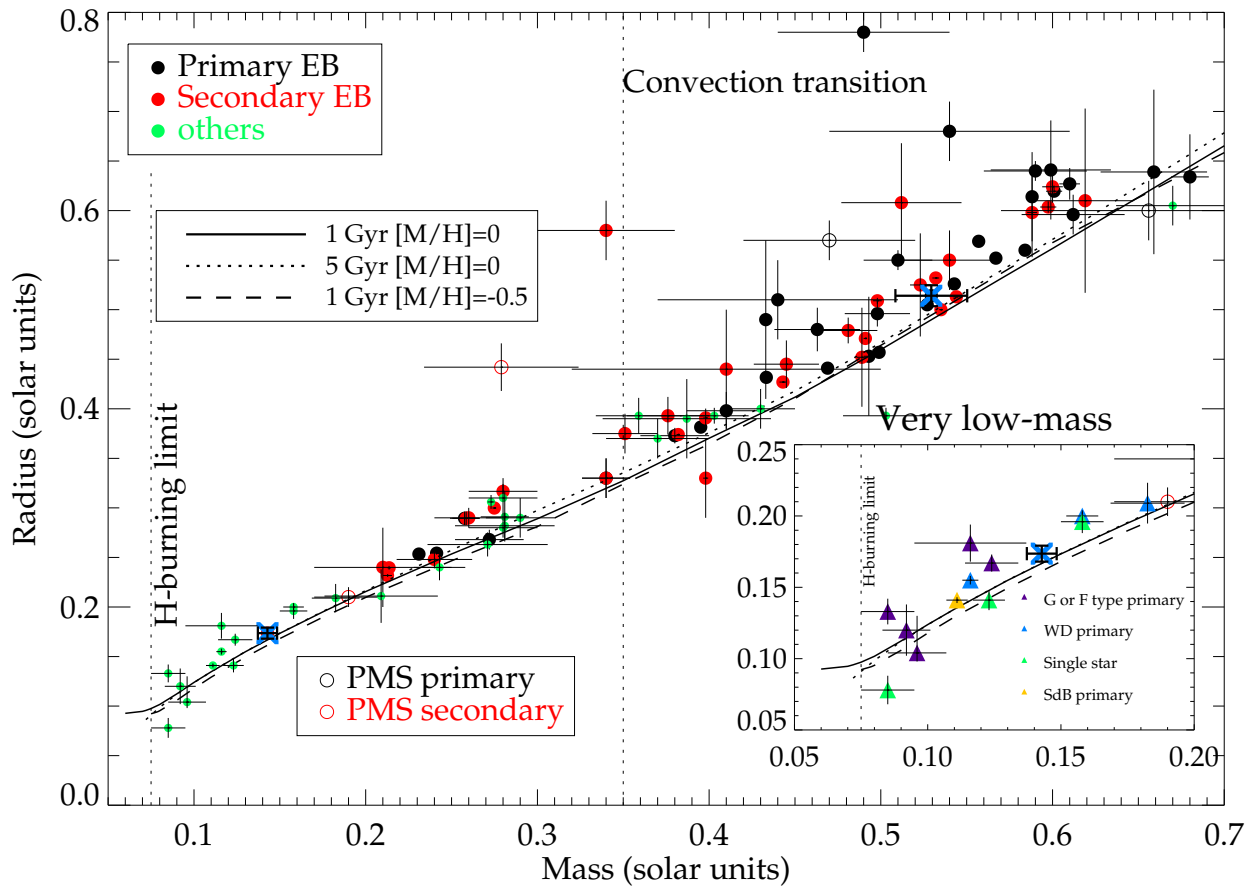}
\caption{\small{The mass-radius diagram for low-mass stars with mass less than 0.7$M_{\odot}$. The vertical dashed line at 0.075$M_{\odot}$ indicates the hydrogen burning limit, whereas the line at $\sim$0.35$M_{\odot}$ represents the proposed transition to fully convective stellar atmospheres (Chabrier \& Baraffe 1997). Black filled dots are the primary stars in double-lined eclipsing M+M-dwarf binaries, whereas red filled dots are the secondaries. The small filled green dots indicate M-dwarfs with masses and radii determined through other methods, which we further examine in the inset for very low mass M-dwarfs (M$<$0.2$M_{\odot}$). The purple, blue, and yellow triangles indicate measurements of M-dwarfs in eclipsing systems with higher mass G or F type primaries, white dwarf (WD) primaries, or subdwarf B-type (SdB) primaries respectively. The green filled triangles are single star measurements using radius measurements from interferometry. }}
\label{Figure9}
\end{figure*}
\begin{figure*}
\includegraphics[width=1.\textwidth]{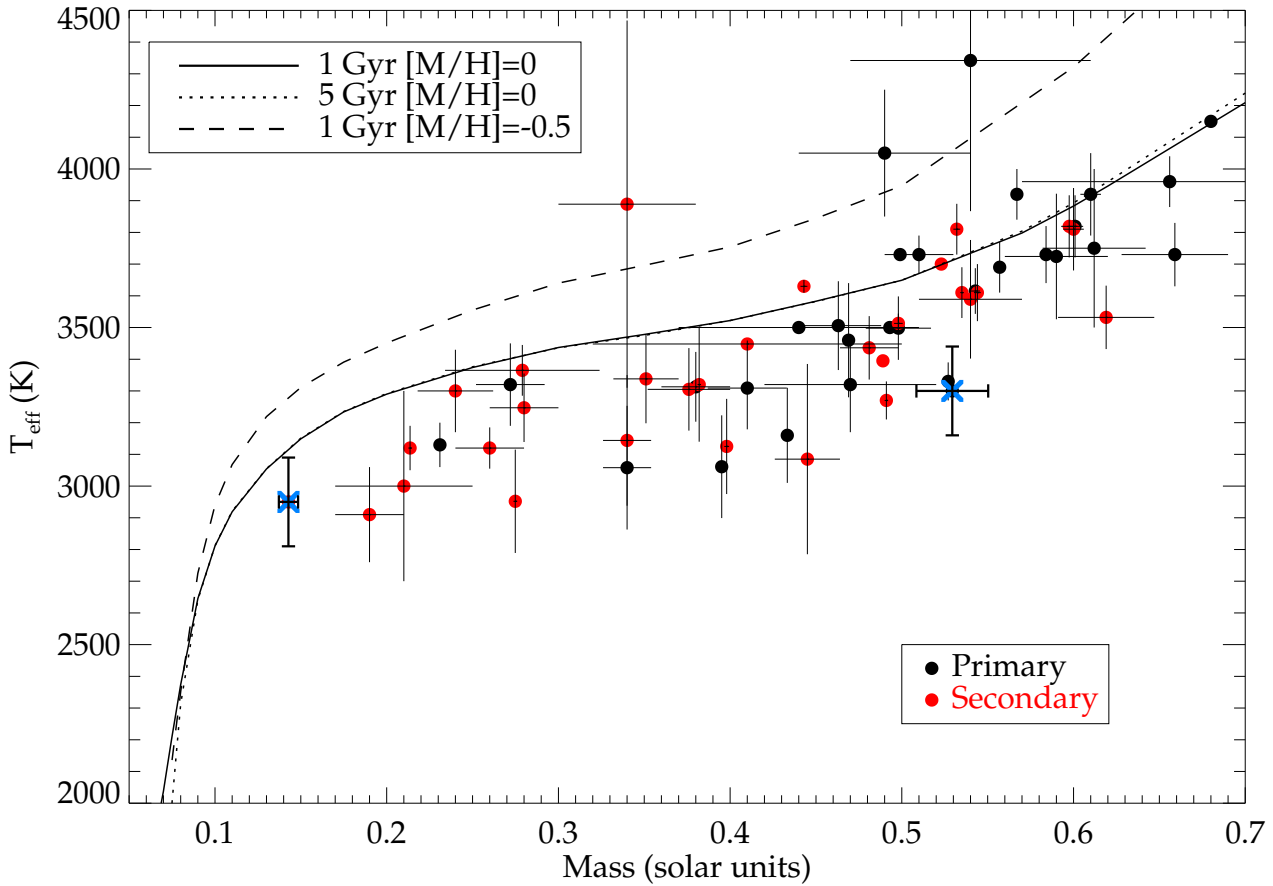}
\caption{\small{Current census of mass, effective temperature measurements for stars with mass less than 0.7$M_{\odot}$. Similar to Figure \ref{Figure9}, filled dots indicate primary (black) and secondary (red) components of double-lined eclipsing binaries. The blue crosses indicate the position of the components of binary WTS 19g-4-02069. The solid black, dotted and dashed curves are Baraffe model predictions for different ages and metallicities. The data for this figure are tabulated in the Appendix.}}
\label{Figure10}
\end{figure*}

In Figures \ref{Figure9} and \ref{Figure10} we present the current sample of mass-radius-temperature measurements for M-dwarfs. In these figures, filled black and red dots indicate dynamically derived model-independent masses and radii for the primaries and secondaries of eclipsing double-lined M+M-dwarf binary systems. Green dots represent other measurements of M-dwarfs which mostly rely on model-dependent assumptions and/or external constraints on the system. These systems include M-dwarfs as secondaries of single line binary systems, tidally interacting white dwarf-M-dwarf systems and single star M-dwarf measurements which often assume an empirical mass-luminosity relation. In the inset of Figure \ref{Figure9} we show the available measurements in the regime $<$0.2$M_{\odot}$, further detailing their origin. There is significant scatter in these data. Four of the data-points are from M-dwarfs orbiting F- or G-type stars, i.e. single line systems, and rely on model-dependent constraints on the properties of the primary and/or assume spin-orbit alignment (Pont et al. 2005, 2006; Beatty 2007). Three of the M-dwarf systems have either a white dwarf primary, which may have had a phase of common envelope evolution or significant mass-transfer and are likely to be tidally interacting, or B-type subdwarf (Parsons et al. 2012;Pyrzas 2012). Three other systems are from interferometric data with directly measured radii, but with estimated masses from a model mass-luminosity relation (e.g. Demory et al. 2009). Finally two data-points are pre-main sequence M-dwarf binaries which have secondaries in the $<$0.2$M_{\odot}$ regime (JW380, Irwin et al. 2007; 2MASSJ04463285, Hebb et al. 2006). The cool companion of WTS 19g-4-02069A is in a special position as it provides a model-independent anchoring point of mass and radius for fully convective low-mass main-sequence M-dwarfs.

The companion is 0.0067$\rm R_{\odot}$ (4.0\%) larger in radius than the 1Gyr solar metallicity Baraffe model, whereas the primary is larger by 0.025$\rm R_{\odot}$ (5.0\%). The radius of the companion can be constrained to 0.0056$R_{\odot}$(3.2\%) at the 1$\sigma$ level, its mass is uncertain by 0.0056$M_{\odot}$(4.0\%), indicating that it is currently outside the $<$3\% range advocated by e.g. Torres (2012) as a stringent constraint on models. We illustrate this by showing three Baraffe (1998) models for: i) 1 Gyr solar metallicity (solid black curve); ii) 5 Gyr solar metallicity (dotted curve) and iii) 1 Gyr metal poor, [M/H]=-0.5 (dashed curve) in Figure \ref{Figure9}. A 3\% accuracy range on the companion mass could be reached if the uncertainty on K$_2$ is pushed down to the $\sim$1 km/s level. It is interesting to note that the effective temperature for the fully-convective secondary is consistent with the 1 Gyr model within the 140 K measurement error, but the partially-convective primary has a $T_{eff}$ lower by $\sim$350 K. Although there is significant scatter in the mass-$T_{eff}$ diagram around 0.5$M_{\odot}$, potentially due to the inhomogeneous set of methods by which temperature is determined, such a difference could be explained by magnetic inhibition theory (e.g. Chabrier et al. 2007) if the primary is an active star and has a significant magnetic field, whereas the secondary may not. 

Five of the seven literature eclipsing binaries with q$<$0.6 are fast rotators ($P<$1 d), and three of these seven are still on pre-main sequence tracks. The primary of main-sequence M-dwarf binary V405 And (Vida et al. 2009; $q\sim$0.429, $P\sim$0.465 d) is inflated with respect to the Baraffe model by 73\%, whereas the secondary is not significantly inflated. This system is much more active than WTS 19g-4-02069, with frequent flaring events. In NSVS-65550671 (Dimitrov \& Kjurkcieva 2010, $q\sim$0.510, $P\sim$0.193 d), both components are inflated ($\sim$17\% for the primary, $\sim$12\% for the secondary). Hebb et al. (2006) present a $q\sim$0.404, $P\sim$0.465 d system in the young open cluster NGC 1647, which has an inflated 0.47$M_{\odot}$ primary and a non-inflated 0.19$M_{\odot}$ secondary. According to Figure 1 from Irwin 2007, 0.5$M_{\odot}$ M-dwarfs reach the main sequence by $\sim$150 Myr, and by $\sim$300 Myr for 0.2$M_{\odot}$, indicating that one or both components of the Hebb et al. binary may be pre-main-sequence. In the pre-main-sequence binary NSVS-06507557 (Cakirh \& Ibanoglu 2010; $q\sim$0.425, $P\sim$0.520 d) the contrary is true when comparing with the models: the secondary is inflated by 35\%, whereas the primary is consistent with the models. One explanation is that in such a young system, the size contraction through gravitational collapse of the secondary may not have progressed as much as that of the higher mass primary, because lower mass stars reach the Zero Age Main Sequence (ZAMS) at later times. Binary WTS 19g-4-02069, for which we find no clear indications of a young age, has a $>$5 times lower rotation rate than these four systems, suggesting that inhibition by stellar rotation is the main reason for the observed difference in radius inflation. 

\subsection{Non-synchronous rotation?}
In Section 4.1 we determine that the WFCAM J-band data exhibit a $\sim$2\% peak-to-peak near-sinusoidal periodicity. The variability is not in phase with the binary eclipses, therefore we attribute it to a rotational modulation in brightness caused by star spots. The H$_{\alpha}$ profile, as observed with ISIS (Section 3.2.1), suggests that the primary star is the main contributor to both the emission of this line and the J-band variability, because the observed radial velocity shift is consistent with the primary. Furthermore, the secondary star contributes only $\sim$7\% of the total J-band emission, therefore requiring spot modulations of order $\sim$0.0162/0.07$\sim$0.23 mag, to account for the variability, which would require a very high spot coverage on the secondary and a magnetically quiet primary. The best-fit period of $\sim$2.563 d, $\sim$0.14 d ($\sim$5\%) longer than the binary orbital period of $\sim$2.44 d, would suggest that the primary star rotates at a subsynchronous rate. This finding is in contrast with current predictions from tidal theory (e.g. Zahn 1977), that suggest orbital synchronisation on timescales of the order 10$^{4-6}$ yr and $\textit{super}$synchronous rotation for young stars. We discuss two possible scenarios that could explain the observations.

In the first scenario, high latitude spots have a longer rotation period than spots near the M-dwarf equator due to significant differential rotation over the stellar surface. This causes the observed rotation rate $\Omega_{obs}=2\pi/P_{obs}$ to be lower than the true equatorial rate $\Omega_{eq}$. On the Sun for example, a spot at 60$^{\circ}$ latitude has a rotation period $\sim$25\% longer than on the equator, and the intermediate latitude $\theta$ is given by the equation $\Omega(\theta)=\Omega_{eq}-d\Omega sin^2(\theta)$, where d$\Omega$ is the difference in rotation between the equator and the poles. Donati et al. (2008) present a spectropolarimetric survey of a small number of single early-type M-dwarfs (ranging from M0 to M3), and detect significant differential rotation in four objects of their sample. For example, OT Ser (M1.5 dwarf; $\sim$3.38 d - $d\Omega/\Omega_{eq}\sim$0.06) and DS Leo (M0 dwarf; $\sim$14 d - $d\Omega/\Omega_{eq}\sim$0.16) show surface rotation variations which are consistent 
with, although somewhat higher than our observations. Barnes et al. (2005) argue that for low-mass stars the surface differential rotation vanishes with increasing convective depth, such that fully convective stars rotate mostly as solid bodies. This is supported by observations from e.g. Donati et al. (2006a) and Morin et al. (2008a,b) who show little to no differential rotation in late-type M-dwarfs. We therefore argue that the variability of WTS 19g-4-02069 could be caused by a large high-contrast stable cool spot complex located near to the rotation pole(s) of the 0.5$M_{\odot}$ primary, where the rotational shear is relatively low. Similar large and stable (nearly) polar spot patterns with (quasi-)sinusoidal light curves are observed on the young $\sim$M2V star AU Mic (see e.g. Rodono et al. 1986). A dichotomy in magnetic field geometry is pointed out by Morin et al. (2010) for stars above and below 0.5$M_{\odot}$, suggesting that the dynamo mechanism for 0.5$M_{\odot}$ stars is rather like that of solar-type stars, where the Coriolis force induced by fast stellar rotation (WTS 19g-4-02069 rotates at $\sim$10$\times$ the solar rotation value) tends to drive up star spots to polar latitudes. The photometric signal in Figure \ref{Figure7} has a relatively high scatter compared to the sine amplitude ($\sim$10.4 mmag scatter for a $\sim$16.2 mmag signal), which may suggest that the size of the region could have fluctuated over the timescale ($\sim$5 years) of our observations. 

In the second scenario, the binary could have gone through an extra phase of spin-down besides tidal dissipation. In the sample of Strassmeier et al. (2012) 74\% of the rapidly-rotating active binary stars are synchronized and in circular orbits, but 26\% (61 systems) are rotating asynchronously of which half have P$_{rot}>$P$_{orb}$, mostly giant stars. It is suggested that a magnetic wind could have applied a braking torque on the stars on the main-sequence, the magnitude of which may depend on stellar mass, interior structure, and activity. This could indicate that the primary and (fully convective) secondary star of WTS 19g-4-02069 may have been braked with different rates, which could be tested by comparing the rotation periods of both stars. Alternatively, the subsynchronous rotation may have been established during the pre-main sequence phase of the binary, through magnetic interaction with a circumbinary disk with a central hole (e.g. Casey et al. 1993). Here, the magnetic field of the primary could have coupled to the (slower) disk motion, slowing down rotation due to angular momentum transfer to the disk. Arguably, this would require WTS 19g-4-02069 to be young, because any asynchronous rotation could be rapidly dissipated through tidal interaction on the main-sequence, however we currently have no observational evidence to suggest a young system.
\section{Conclusion}
In this work we present the discovery of a highly unequal-mass eclipsing M-dwarf binary (q=0.27) with P=2.44 d, using the high-precision near-infrared time-series of the WFCAM Transit Survey. We find stellar masses $M_1$=0.53 ($\pm$0.02)$M_{\odot}$ and $M_2$=0.143 ($\pm$0.006)$M_{\odot}$, and radii of $R_1$=0.51 ($\pm$0.01)$R_{\odot}$ and $R_2$=0.174 ($\pm$0.006)$R_{\odot}$. The companion star is therefore in a sparsely sampled and important M-dwarf mass-regime for studies of Earth-like exoplanets which require accurate calibration of their host star radius and mass. We suggest that the low mass-ratio of our binary may be explained by the different accretion physics such as recently proposed by 2-D warm grid-based (Ochi et al. 2005, de Val-Borro et al. 2011) or magnetic field braking simulations (Zhao \& Li 2012), which suggest that short period low-q pairs may be abundant around primaries within a large mass range. Alternatively, the binary may have been isolated from the birth environment early on through ejection or rapid gas removal. Since both stars share the same metallicity and age and straddle the theoretical dividing line between fully and partially convective atmospheres, a comparison can be made to model stellar atmospheres with the same isochrone over a wide span of masses. We find that both stars have slightly inflated radii compared to 1 Gyr model predictions for their mass, but we argue that future work will be required to quantify the effects of star spots on the light curve solution. The effective temperature of the secondary is consistent with theoretical models, but for the primary it is lower by $\sim$350 K, which could be explained by magnetic inhibition theory (e.g. Chabrier et al. 2007) if the primary is an active star and has a significant magnetic field, whereas the secondary may not. The detection of a 2.56 d $\sim$2\% signal in the WFCAM light curve is attributed to subsynchronous rotation of a relatively stable star-spot complex at high latitude on the magnetically active primary, suggesting that its dynamo is more like solar-type stars, with the Coriolis force driving up star spots to polar latitudes.
\section{Acknowledgements}
We thank the anonymous referee for comments that led to the improvement of this manuscipt. SVN, JLB, IAGS,SH, DJP and EM have received support from RoPACS during this research, and BS, GK are supported by RoPACS, a Marie Curie Initial Training Network funded by the European Commissions Seventh Framework Programme. The United Kingdom Infrared Telescope is operated by the Joint Astronomy Centre on behalf of the Science and Technology Facilities Council of the U.K. This article is based on observations made with the INT and WHT, operated on the island of La Palma by the ING in the Spanish Observatorio del Roque de los Muchachos, La Palma. This research has benefitted from the M, L, and T dwarf compendium housed at DwarfArchives.org and maintained by Chris Gelino, Davy Kirkpatrick, and Adam Burgasser. This research uses products from SDSS DR7. Funding for the SDSS and SDSS-II has been provided by the Alfred P. Sloan Fundation, the U.S. Department of Energy, the National Aeronautics and Space Administration, the Japanese Monbukagakusho, the Max Planck Society, and the Higher Education Funding Counsil for England. This publication also makes use of data products from the Two Micron All Sky Survey, which is a joint project of the University of Massachusetts and the Infrared Processing and Analysis Center/California Institue of Technology, funded by the National Aeronautics and Space Administration and the National Science Foundation.

\clearpage
\newpage

\section*{Appendix}
\setlength{\tabcolsep}{0.2pt}
\begin{table*}
\scriptsize \begin{tabular*}{1.0\textwidth}
{@{\extracolsep{\fill}}llllllllllll}
   \hline \hline
Name		 	&Ref.			        &$\rm M_1$	 &$\rm M_2$	 &$\rm R_1$	  &$\rm R_2$	  &$\rm T_1$ &$\rm T_2$ &$\rm P$ &$\rm type$ &$\rm (\frac{T_2}{T_1})$ &$\rm q$\\
			&				&$\rm (M_{\odot})$ &$\rm (M_{\odot})$ &$\rm (R_{\odot})$ &$\rm (R_{\odot})$		  &$\rm(K)$	     &$\rm(K)$	&$\rm (d)$	 &           &     		      &\\ 		\hline
OGLEBW3V38              &(1)                            &0.44(0.07)      &0.41(0.09)     &0.51(0.04)      &0.44(0.06)     &3500    &3448(11)  &0.198 &SB2  &0.985    &0.950\\
2MASSJ0154              &(2)                            &0.659(0.031)    &0.619(0.028)   &0.639(0.083)    &0.610(0.093)   &3730(100) &3532(100) &2.639 &SB2  &0.947   &0.939\\
NSVS-6550671            &(3)                            &0.510(0.02)     &0.260(0.02)    &0.550(0.01)     &0.290(0.01)    &3730(60)  &3120(65)  &0.193 &SB2  &0.836   &0.510\\
2MASSJ04463285          &(4)                            &0.470(0.05)     &0.190(0.02)    &0.570(0.02)     &0.210(0.01)    &3320(150) &2910(150) &0.630 &SB2  &0.877   &0.404\\
NSVS-06507557           &(5)                            &0.656(0.086)    &0.279(0.045)   &0.600(0.03)     &0.442(0.024)   &3960(80)  &3365(80)  &0.520 &SB2  &0.850   &0.425\\
NSVS-01031772           &(6)                            &0.5428(0.0027)  &0.498(0.0025)  &0.526(0.0028)   &0.509(0.003)   &3615(72)  &3513(31)  &0.368 &SB2  &0.972   &0.917\\
GJ3236           	&(8)             	        &0.38(0.02)      &0.28(0.02)     &0.3729(0.0078)  &0.3167(0.0075) &3313(110) &3247(108) &0.771 &SB2  &0.98    &0.737\\
CUCnc                   &(7)                            &0.4333(0.0017)  &0.398(0.0014)  &0.4317(0.0052)  &0.3908(0.0094) &3160(150) &3125(150) &2.77  &SB2  &0.989   &0.919\\
SDSS-MEB-1              &(9)                            &0.272(0.02)     &0.24(0.022)    &0.268(0.01)     &0.248(0.009)   &3320(130) &3300(130) &0.407 &SB2  &0.99    &0.880\\
V405And                 &(10)                           &0.49(0.05)      &0.21(0.04)     &0.78(0.02)      &0.24(0.04)     &4050(200) &3000(300) &0.465 &SB2  &0.741   &0.429\\
ASAJ011328-3821.1       &(11)                           &0.612(0.03)     &0.445(0.019)   &0.596(0.02)     &0.445(0.024)   &3750(250) &3085(300) &0.445 &SB2  &0.822   &0.727\\
LP-133-373              &(12)                           &0.34(0.014)     &0.34(0.014)    &0.33(0.02)      &0.33(0.02)     &3058(195) &3144(206) &1.63  &SB2  &0.973   &1.000\\ 
T-Lyr1-17236            &(13)                           &0.680(0.011)    &0.523(0.006)   &0.634(0.043)    &0.525(0.052)   &4150    &3700   &8.43  &SB2  &0.892   &0.769\\
CMDra            	&(14)                           &0.231(0.001)    &0.2136(0.001)  &0.2534(0.0019)  &0.2396(0.002)  &3130(70)  &3120(70)  &1.27  &SB2  &0.997   &0.925\\
LSPMJ1112+7626          &(15)                           &0.3951(0.0022)  &0.2749(0.0011) &0.3860(0.005)      &0.2978(0.005)     &3061(162) &2952(163) &41.03 &SB2  &0.964   &0.696\\
1RXSJ154727             &(16)                           &0.2576(0.0085)  &0.2585(0.008)  &0.2895(0.0068)  &0.2895(0.0068) &--  &--  &3.55  &SB2  &--  &0.997\\
WTS19b-2-01387          &(17)                           &0.498(0.019)    &0.481(0.017)   &0.496(0.013)    &0.479(0.013)   &3498(100) &3436(100) &1.499 &SB2  &0.982   &0.966\\
WTS19c-3-01405          &(17)                           &0.410(0.023)    &0.376(0.024)   &0.398(0.019)    &0.393(0.019)   &3309(130) &3305(130) &4.939 &SB2  &0.999   &0.917\\
WTS19e-3-08413          &(17)                           &0.463(0.025)    &0.351(0.019)   &0.480(0.022)    &0.375(0.02)    &3506(140) &3338(140) &1.673 &SB2  &0.952   &0.758\\
JW380	                &(18)                           &0.26(0.02)	 &0.15(0.01)     &1.19(0.11)      &0.90(0.10)     &--  &--  &5.3   &SB2  &--  &0.58\\
KOI126BC                &(19)                           &0.2413(0.003)   &0.2127(0.0026) &0.2543(0.0014)  &0.2318(0.0013) &--  &--  &1.767 &SB2  &--  &0.881\\                   
MG1-646680              &(20)                           &0.499(0.002)    &0.443(0.002)   &0.457(0.005)    &0.427(0.004)   &3730(20) &3630(20) &1.638 &SB2  &0.973    &0.888\\
MG1-78457               &(20)                           &0.527(0.002)    &0.491(0.001)   &0.505(0.0075)   &0.471(0.008)   &3330(60) &3270(60) &1.586 &SB2  &0.982    &0.932\\
MG1-116309              &(20)                           &0.567(0.002)    &0.532(0.002)   &0.552(0.0085)   &0.532(0.006)   &3920(80) &3810(80) &0.827 &SB2  &0.972    &0.938\\   
MG1-1819499             &(20)                           &0.557(0.001)    &0.535(0.001)   &0.569(0.0022)   &0.500(0.0085)  &3690(80) &3610(80) &0.630 &SB2  &0.978    &0.961\\
MG1-506664              &(20)                           &0.584(0.002)    &0.544(0.002)   &0.560(0.0025)   &0.513(0.0055)  &3730(90) &3610(90) &1.548 &SB2  &0.968    &0.932\\
MG1-2056316             &(20)                           &0.469(0.002)    &0.382(0.001)   &0.441(0.002)    &0.374(0.002)   &3460(180) &3320(180) &1.723  &SB2  &0.960    &0.814\\ 
SDSSJ001641-000925      &(21)				&0.54(0.07)      &0.34(0.04)     &0.68(0.03)      &0.58(0.03)     &4342(475) &3889(579) &0.199  &SB2  &0.896    &0.630\\
TrES-Her0-07621         &(22)                           &0.493(0.003)    &0.489(0.003)   &0.453(0.06)     &0.452(0.05)    &3500      &3395      &1.137  &SB2  &0.97     &0.992\\
BD-225866Aa             &(23)                           &0.5881(0.0029)  &0.5881(0.0029) &0.614(0.045)    &0.598(0.045)   &--        &--        &2.211  &SB2  &--       &1.000\\
UNSW2A                  &(24)                           &0.599(0.035)    &0.512(0.035)   &0.641(0.05)     &0.608(0.06)    &--        &--        &2.144  &SB2  &--       &0.855\\
HIP96515Aa              &(25)                           &0.59(0.03)      &0.54(0.03)     &0.64(0.01)      &0.55(0.03)     &3724(198) &3589(187) &2.346  &SB2  &0.964    &0.915\\
YYGem                   &(26)                           &0.6009(0.0047)   &0.5975(0.0047)  &0.6196(0.0057)    &0.6036(0.0057)   &3819(98)  &3819(98)  &0.814  &SB2  &1.000    &0.994\\
GUBoo                  &(27)                           &0.610(0.006)    &0.600(0.006)   &0.627(0.016)    &0.624(0.016)   &3920(130) &3810(130) &0.492  &SB2  &0.972    &0.984\\
& & & & & & & & & & &\\
HAT-TR-205-013	 	&(28)          		        &      &0.124(0.01)    &      &0.167(0.006)	  &          &          &      &SB1  &        &\\
OGLE-TR-5B       	&(29)                           &      &0.271(0.035)   &      &0.263(0.012)       &          &          &         &      SB1  &        &\\
OGLE-TR-6B       	&(29)                           &      &0.359(0.025)    &     &0.393(0.018)       &          &          &          &     SB1  &        &\\
OGLE-TR-7B       	&(29)                           &      &0.281(0.029)    &     &0.282(0.013)       &          &          &          &     SB1  &        &\\  
OGLE-TR-123B     	&(30)                           &      &0.085(0.01)     &     &0.133(0.009)       &          &          &          &     SB1  &        &\\
OGLE-TR-122B     	&(31)                           &      &0.092(0.009)    &     &0.120(0.018)       &          &          &          &     SB1  &        &\\
OGLE-TR-106B     	&(31)                           &      &0.116(0.021)    &     &0.181(0.013)       &          &          &          &     SB1  &        &\\
OGLE-TR-125B     	&(31)                           &      &0.209(0.033)    &     &0.211(0.027)       &          &          &          &    SB1  &        &\\
OGLE-TR-78B      	&(31)                           &      &0.243(0.015)    &     &0.24(0.013)        &          &          &          &     SB1  &        &\\
OGLE-TR-18B      	&(32)                           &      &0.387(0.049)    &     &0.390(0.040)       &          &          &          &    SB1  &        &\\
T-Aur0-13378       &(33)                           &      &0.37(0.03)    &     &0.37(0.02)       &          &          &          &    SB1  &        &\\
T-Boo0-0080       &(33)                           &      &0.28(0.02)    &     &0.31(0.02)       &          &          &          &    SB1  &        &\\
T-Lyr1-01662       &(33)                           &      &0.28(0.02)    &     &0.28(0.01)       &          &          &          &    SB1  &        &\\
T-Lyr0-08070       &(33)                           &      &0.29(0.02)    &     &0.29(0.02)       &          &          &          &    SB1  &        &\\
T-Cyg1-01385       &(33)                           &      &0.43(0.02)    &     &0.40(0.02)       &          &          &          &    SB1  &        &\\
CoRoT101186644       &(34)                           &      &0.096(0.011)    &     &0.104$^{+0.026}_{-0.006}$       &          &          &          &    SB1  &        &\\
GJ191                   &(35)                             &0.281(0.014)    &single         &0.291(0.025)    &single         &3570(156)          &          &     &single &       &\\
GJ699                   &(35)                             &0.158(0.008)    &single         &0.196(0.008)    &single         &3163(65)          &          &     &single &       &\\
GJ551                   &(35)                           &0.123(0.006)    &single         &0.141(0.007)    &single         &3042(117)          &          &     &single &       &\\
GJ887                   &(35)                           &0.503(0.025)    &single         &0.393(0.008)    &single         &          &          &     &single &       &\\
GJ411                   &(35)                             &0.403(0.02)     &single         &0.393(0.008)    &single         &3570(42)          &          &     &single &       &\\
GJ380                   &(35)                             &0.670(0.033)    &single         &0.605(0.02)     &single         &          &          &     &single &       &\\
RRCae-B                 &(36)                             &WD              &0.1825(0.0139) &WD              &0.209(0.0143)  &          &3100(100)          &     &WD     &       &\\
NNSer-B                 &(37)                             &SdB             &0.111(0.004)   &SdB             &0.141(0.002)   &          &          &     &SdB    &       &\\     
GKVir                   &(38)                           &WD              &0.116(0.003)   &WD              &0.155(0.003)   &          &          &     &WD     &       &\\
SDSSJ1212 0123          &(39)                           &WD              &0.273(0.002)   &WD              &0.306(0.007)   &          &          &     &WD     &       &\\
SDSSJ1210+3347          &(39)                           &WD              &0.158(0.006)   &WD              &0.200(0.003)   &          &          &     &WD     &       &\\
\hline
\end{tabular*}
\label{Table6}
\caption{{\small{Literature values for the M-dwarf systems used in Figure \ref{Figure8}, \ref{Figure9} and \ref{Figure10}. References: (1) Maceroni \& Montalban (2004), (2) Becker et al. (2008), (3) Dimitrov \& Kjurkchieva (2010), (4) Hebb et al. (2006), (5) Cakirh \& Ibanoglu (2010), (6) Lopez-Morales et al. (2006), (7) Ribas (2003), (8) Irwin et al. (2009), (9) Blake et al. (2008), (10) Vida et al. (2009), (11) Helminiak et al. (2012), (12) Vaccaro et al. (2007), (13) Devor et al. (2008), (14) Morales et al. (2009), (15) Irwin et al. (2011), (16) Hartman et al. (2011), (17) Birkby et al. (2012), (18) Irwin et al. (2007), (19) Carter et al. (2011), (20) Kraus et al. (2011), (21) Davenport et al. (2012), (22) Creevey et al. (2005), (23) Shkolnik et al. (2010), (24) Young et al. (2006), (25) Huelamo et al. (2009), (26) Torres \& Ribas (2002), (27) L{\'o}pez-Morales \& Ribas (2005), (28) Beatty (2007), (29) Bouchy et al. (2005), (30) Pont et al. (2006), (31) Pont et al. (2005), (32) Bouchy et al. (2005), (33) Fernandez et al. (2009), (34) Tal-Or et al. (2013), (35) Segransan et al. (2002), (36) Maxted et al. (2007), (37) Parsons et al. (2010), (38) Parsons et al. (2012), (39) Pyrzas et al. (2012).}}}
\end{table*}

\cleardoublepage

\end{document}